\documentclass[lettersize,journal]{IEEEtran}
\usepackage{tikz}

\usepackage[ruled,lined,linesnumbered,commentsnumbered,longend]{algorithm2e}
\usepackage{tikz}
\usepackage{amsmath}
\usepackage{hyperref}
\usepackage{graphicx}
\usepackage{subcaption}
\usepackage{amssymb}
\usepackage{booktabs} 
\usepackage{mathrsfs}
\usepackage{xcolor}
\usepackage{amsthm}
\newtheorem{theorem}{Theorem}
\newcommand{\equalcontribution}{\textsuperscript{*}}
\newcommand{\correspondingauthor}{\textsuperscript{\dag}}

\begin{document}

\title{BriDe Arbitrager: Enhancing Arbitrage in Ethereum 2.0  via Bribery-enabled Delayed Block Production}

\author{
Hulin Yang\equalcontribution, Mingzhe Li\equalcontribution,~\IEEEmembership{Member,~IEEE},~Jin~Zhang\correspondingauthor,~\IEEEmembership{Member,~IEEE}, Alia Asheralieva, \\
Qingsong Wei,~\IEEEmembership{Senior Member,~IEEE},
        and Siow Mong Rick Goh,~\IEEEmembership{Senior Member,~IEEE}

\IEEEcompsocitemizethanks{\IEEEcompsocthanksitem H. Yang is with the Research Institute of Trustworthy Autonomous Systems and Department of Computer Science and Engineering, Southern University of Science and Technology, Shenzhen, China, 518055 (email: yanghl2021@mail.sustech.edu.cn).
\IEEEcompsocthanksitem M. Li is with the Institute of High Performance Computing (IHPC), Agency for Science, Technology and Research (A*STAR), Singapore (email: mlibn@connect.ust.hk, Li\_Mingzhe@ihpc.a-star.edu.sg).
\IEEEcompsocthanksitem J. Zhang is with the Research Institute of Trustworthy Autonomous Systems and Department of Computer Science and Engineering, Southern University of Science and Technology, Shenzhen, China, 518055 (email: zhangj4@sustech.edu.cn).
\IEEEcompsocthanksitem A. Asheralieva is with the School of Computer Science, Loughborough University, UK (e-mail: aasheralieva@gmail.com).
\IEEEcompsocthanksitem Q. Wei and S. Goh are with the Institute of High Performance Computing (IHPC), Agency for Science, Technology and Research (A*STAR), Singapore (email: wei\_qingsong@ihpc.a-star.edu.sg, gohsm@ihpc.a-star.edu.sg).
\IEEEcompsocthanksitem \equalcontribution H. Yang and M. Li are the co-first author.
\IEEEcompsocthanksitem \correspondingauthor J. Zhang is the corresponding author.
}
\thanks{Copyright (c) 20xx IEEE. Personal use of this material is permitted. However, permission to use this material for any other purposes must be obtained from the IEEE by sending a request to pubs-permissions@ieee.org.}
}

\markboth{Journal of \LaTeX\ Class Files,~Vol.~14, No.~8, August~2024}%
{Shell \MakeLowercase{\textit{et al.}}: A Sample Article Using IEEEtran.cls for IEEE Journals}

\maketitle

\begin{abstract}

The advent of Ethereum 2.0 has introduced significant changes, particularly the shift to Proof-of-Stake consensus. This change presents new opportunities and challenges for arbitrage. 
Amidst these changes, we introduce BriDe Arbitrager, a novel tool designed for Ethereum 2.0 that leverages \underline{Bri}bery-driven attacks to \underline{De}lay block production and increase arbitrage gains. The main idea is to allow malicious proposers to 
delay block production
by bribing validators/proposers, thereby gaining more time to identify arbitrage opportunities. 
Through analysing the bribery process, we design an adaptive bribery strategy.
Additionally, we propose a Delayed Transaction Ordering Algorithm to leverage the delayed time to amplify arbitrage profits for malicious proposers. To ensure fairness and automate the bribery process, we design and implement a bribery smart contract and a bribery client. As a result, BriDe Arbitrager enables adversaries controlling a limited ($<1/4$) fraction of the voting powers to delay block production via bribery and arbitrage more profit. Extensive experimental results based on Ethereum historical transactions demonstrate that BriDe Arbitrager yields an average of 8.66\,ETH (16,442.23\,USD) daily profits. Furthermore, our approach does not trigger any slashing mechanisms and remains effective even under Proposer Builder Separation and other potential mechanisms will be adopted by Ethereum.
\end{abstract}


\section{Introduction}
With the popularity of Decentralized Finance (DeFi)~\cite{zetzsche2020decentralized} -- a decentralized financial system that works automatically on the blockchain, 
users of blockchain can extract 
Maximal/Miner Extractable Value (MEV)~\cite{daian2020flash} from DeFi through various approaches~\cite{zhou2021high, zhou2021just, qin2021empirical, qin2022quantifying}.
Among all kinds of MEV extractions, \textbf{\textit{Arbitrage}} (i.e., simultaneously buying and selling an asset in different markets to profit from a price difference) is one of the primary means, accounting for 42\% of all MEV profits~\cite{qin2022quantifying}.
Previously, arbitrage has been investigated during the era of Ethereum 1.0~\cite{daian2020flash, qin2022quantifying, zhou2021just, qin2021attacking}. 
However, Ethereum, being the project profoundly influenced by arbitrage, transitioned to 2.0 on September 15, 2022. 
As a result, investigating arbitrage related concerns on Ethereum 2.0 has substantial merit.

A critical change of Ethereum 2.0 is that its consensus mechanism has changed from proof-of-work (PoW)~\cite{nakamoto2008bitcoin} to proof-of-stake (PoS)~\cite{king2012ppcoin}. 
During the era of Ethereum 1.0, miners generate blocks through competition in computational power~\cite{buterin2014next}. 
While in Ethereum 2.0, the block production process evolves into a system divided into fixed-length slots. 
For each slot, a proposer is selected randomly to produce a block, 
followed by the validators of the consensus committee casting their votes on the proposed block~\cite{buterin2020combining}. 
Ultimately, the network adopts the chain with the maximum vote weight as the primary chain for finality. 
This paradigm shift presents new opportunities and challenges for arbitrage.

The fresh opportunity we observe is that, since there is no proposer competition in Ethereum 2.0, a proposer could potentially \emph{\textbf{delay block production}}.
Therefore, the proposer might \emph{\textbf{gain more arbitrage profits}} as it has more time to observe additional information.
However, it is challenging to achieve this.
Some existing works attempted to propose ex ante reorganization attacks~\cite{schwarz2022three, neuder2021low} that allow proposers to delay block release.
However, some Ethereum community's proposals, such as the proposer boosting mechanism~\cite{proposerboosting}, 
have been able to make such attacks (e.g., delaying block production) difficult to launch.
In the proposer boosting mechanism, the newly produced block of the current slot will be given a temporary weight of $1/4$ of the current slot's total voting weights. 
Validators are encouraged to vote for this block, giving it an advantage.
With the proposer boosting mechanism, a malicious proposer needs to control more than $1/4$ of the voting powers to delay block production, which is hard to achieve successfully. 
Otherwise, the delayed produced blocks will be discarded~\cite{proposerboosting}.

\vspace{3pt}
\noindent
\textbf{Bribery-Enabled Delayed Arbitrage. }
How can a proposer delay block production to increase its arbitrage profit while holding a limited fraction of voting powers ($<1/4$)?
In this work, we propose a bribery attack to enable delayed block production for proposers controlling a limited fraction of voting powers, thus enhancing arbitrage profits. 
The central idea is that a \emph{malicious proposer (briber)} delays block production by bribing \emph{other validators or proposers (bribees)} from different slots, enabling the block it produces to attain a higher voting weight and be preserved by the fork choice rule~\cite{buterin2020combining}. 

The use of bribery attacks in Ethereum 2.0 is \textbf{\emph{justifiable}}.
First, blockchain nodes are driven by incentives~\cite{bonneau2016buy}; such rational nodes have sufficient motivation to accept bribes to augment their earnings. 
Second, although the malicious proposer incurs bribery costs as the briber, the extra time they secure for arbitrage through bribery can lead to additional profits. 
Last, 
unlike certain previous attacks~\cite{neu2022two}, our bribery attack does not trigger any slashing mechanisms.

\vspace{3pt}
\noindent
\textbf{Challenges.}
To successfully execute bribery-enabled delayed arbitrage, several challenges need to be addressed.
(I) \emph{\textbf{How to design a wise bribery strategy for proposers which only control a limited fraction of voting powers?}}
We need to decide the appropriate delay time, bribery object and bribery fee to successfully delay block production at a low cost. (II) \emph{\textbf{How to fully leverage the delayed time to amplify the arbitrage profit to achieve high revenue}?}
The proposed bribery attack allows malicious proposers to delay block production and gain more time to observe transactions and arbitrage opportunities.
Therefore, we need to design an algorithm to extract more profitable arbitrage opportunities based on a longer observing time. (III) \emph{\textbf{How to execute bribery and arbitrage fairly and automatically?}} A briber might renege on paying after the requested actions are performed, while a bribee might accept the bribe but fail to carry out the corresponding actions. 
This could \emph{compromise the fairness} of bribery.
Moreover, bribers should automatically decide whether to launch the bribery attack while bribees should automatically decide whether to accept the bribe, for usability and feasibility.

\vspace{3pt}
\noindent
\textbf{Bribery Attack Formalization.}
To address the first challenge, we formalize the entire bribery process. 
We investigate the selection of potential delay time, bribery object and bribery cost.
Based on the analysis, delaying to the next slot to produce the block is the optimal delay time. 
As for the bribery object selection, the malicious proposer decides whether to bribe the validators of the next slot to vote for the delayed block or to bribe the proposer to give up the block proposing opportunity. According to the number of bribable validators, the malicious proposer selects the bribery object with minimal bribery fee.
Under our strategy, both malicious proposer (briber) and rational validator/proposer (bribees) can gain positive profits.

\vspace{3pt}
\noindent
\textbf{Delayed Transaction Ordering Algorithm.}
To address the second challenge, we propose Delayed Transaction Ordering Algorithm (DTOA): 
a cyclic arbitrage algorithm based on delayed block production and transaction ordering to magnify the arbitrage profits.
First, among numerous liquidity pools, we select some liquidity pools with higher transaction frequency and lower liquidity, where more likely to have transactions that can cause greater price fluctuation.
Second, delayed block production allows us to observe more transactions, which results in more arbitrage opportunities.
However, existing cyclic arbitrage algorithms based on negative cycle detection~\cite{zhou2021just} offer unsatisfactory profits, because they focus on finding arbitrage opportunities faster and miss higher profit arbitrage opportunities.
Therefore, we propose an arbitrage algorithm based on \emph{Depth-First Search (DFS) cycle detection} to find the best arbitrage opportunity for each round.
Last, we select some appropriate accumulated transactions in the mempool and insert them ahead of our arbitrage transaction (transition ordering), so that other transactions can amplify price fluctuations and \emph{increase the profits} for our arbitrage transaction.


\vspace{3pt}
\noindent
\textbf{Fair and Automated Bribery.}
To address the third challenge, we design a \emph{bribery smart contract} to enable the briber and bribee to complete the bribery process without mutual trust, and a \emph{bribery client} to automate the execution of bribery and arbitrage.
For the bribery contract, 
the briber uses it to verify whether the bribees perform the required actions and provide the bribery fee.
The bribees use the bribery contract to withdraw the bribery fee after they carry out the corresponding actions. Thus, the entire bribery process is controlled through smart contract so that no one can break the rules of bribery. 
This ensures the fairness during the bribery.
For the bribery client, it can help the bribers automatically decide whether to call the bribery contract to launch a bribery attack and provide the bribery fee, according to the arbitrage opportunity and blockchain state. 
The bribery client can also assist the bribees to automatically parse the current block to see if there is a bribery request, and decide whether to accept the bribery and perform required actions.

The main contributions of the paper are as follows:
\begin{itemize}
\item We propose \emph{\textbf{BriDe Arbitrager}}, a tool to boost arbitrage in Ethereum 2.0 via \underline{Bri}bery-enabled \underline{De}layed block production. 
We derive the bribery strategy for adversaries (malicious proposers) \emph{controlling a limited fraction of voting powers} to delay block production and increase arbitrage profits.
\item We design an arbitrage algorithm DTOA based on transaction ordering to amplify bribers' arbitrage profit.
\item We design a bribery smart contract to ensure fairness during bribery.
We then design a bribery client to automate the bribery process to enhance usability. 
\item 
Extensive experimental results show that BriDe Arbitrager finds a total of 12,230 strategy sets yielding profits of 137.83\,ETH (261,794.30\,USD) from Ethereum block 12,000,000 to block 12,100,000. Results and analysis also show that BriDe Arbitrager can still be successfully executed and offer substantial returns even under Proposer Builder Separation~\cite{pbs2021} (PBS) and Secret non-Single Leader Election~\cite{ssle2022} (SSLE).
\end{itemize}

\vspace{-3pt}
\section{Background, Related Works and Motivation}

\subsection{Decentralized Finance}
Decentralized Finance (DeFi) is a blockchain and smart contract-based financial system that allows users to take traditional financial actions such as lending, asset trading, and insurance without trusting a third party. DeFi has attracted significant attention. At the time of writing, the DeFi ecosystem boasts a total market cap of over 66B\,USD~\cite{defi2023}. One of the most well-known applications of DeFi is the Decentralized Exchange (DEX).

\noindent
\textbf{Automated Market Maker.} To adapt to the limited throughput of the blockchain, most current DEXs exist in the form of the automated market maker (AMM)~\cite{adams2021uniswap, hertzog2017bancor, sushiswap, 1inch}. AMM stores assets deposited by liquidity providers in the corresponding liquidity pools, which determine asset prices according to a predefined function. The liquidity taker swaps the desired assets directly with the liquidity pool. 
Among the various types of AMM, the constant product AMM model is the most common, with more than 66\% of the total~\cite{zhou2021just}. 
The constant product AMM model is based on the constant product formula, also known as the Uniswap formula~\cite{adams2021uniswap}, which requires that the product of the number of assets in the liquidity pool remains constant before and after the transaction. 
This work focuses on searching for arbitrage opportunities among constant product AMMs. 

\vspace{-3pt}
\subsection{MEV and Arbitrage}
The concept of MEV was first introduced by Daian et al.~\cite{daian2020flash}. It is the revenue that a miner/validator can earn by including, excluding, and ordering transactions in a block. 
Previous works~\cite{daian2020flash, zhou2021just, zhou2021high, qin2021empirical, qin2022quantifying, qin2023blockchain, qin2021attacking, yaish2022squeezing, heimbach2022eliminating, wang2022cyclic} have investigated MEV in Ethereum 1.0 (PoW) extensively. 
Qin et al.~\cite{qin2022quantifying} show a whopping 540.54M USD of MEV extracted by traders over a nine month period,
with arbitrage accounting for 42\% of the total. 

As the most dominant MEV extraction method, the core idea of arbitrage is to buy at low prices and sell at high prices. 
High-frequency traders monitor the prices of crypto assets on different exchanges in real time. Whenever they find it profitable (i.e., there is a price gap for a certain asset on different exchanges), they send out arbitrage transactions into the blockchain network. 
Zhou et.al.~\cite{zhou2021just} proposed to detect arbitrage opportunities using a negative cycle detection algorithm, which usually misses optimal arbitrage opportunities and takes a lot of time to parameterize arbitrage opportunities. 
McLaughlin et al.~\cite{mclaughlin2023large} give a formula for the arbitrage return and optimize the parameters using a binary search. However, neither of these works attempts to increase the arbitrage revenue using transaction ordering.


Ethereum, the victim of MEV, upgraded its consensus protocol to PoS on September 15, 2022, which brings new opportunities and challenges for MEV extraction arbitrage. To the best of our knowledge, arbitrage under Ethereum 2.0 has not been well-studied. Therefore, this work aims to explore the potential arbitrage opportunities in Ethereum 2.0.

\vspace{-3pt}
\subsection{Ethereum 2.0}
\vspace{-3pt}
In contrast to Ethereum 1.0, where miners used computing power to prove their contributions to earn rewards, Ethereum 2.0 uses the PoS-based consensus protocol Gasper~\cite{buterin2020combining}, which combines the Friendly Finality Gadget Casper FFG~\cite{buterin2017casper} with the Latest Message Driven (LMD) Greediest Heaviest Observed SubTree (GHOST) fork choice rule. 
The consensus process is divided into epochs, where each contains 32 slots. Each slot lasts 12 seconds and is divided into three phases.


Validators are randomly and uniformly divided into 32 committees, each responsible for the consensus of a specific slot. Validators use the LMD GHOST rule to determine the canonical chain head block, i.e., the leaf block of the branch with the heaviest historical vote weight (GHOST).
The total weight of the votes (attestations) collected by the blocks within the branch is calculated for each branch. 
Only each validator's most recent vote (LMD) is considered, and any equivocal votes are ignored. 

One validator is randomly selected as a block proposer per slot from the committee of that slot based on the amount of staked Ether. The proposer is required to determine the canonical chain head block at the beginning of the first phase (0th seconds of the slot) using the LMD GHOST rule and connect the new block to it. The other validators in the committee are required to vote for the canonical chain head block under their views at the beginning of the second phase (4th seconds of the slot). 
Validators with correct votes receive the corresponding attestation reward. The proposer is expected to include the votes of other validators and the output of the sync committee (which is responsible for signing blocks that support lightweight clients) in the block it produces~\cite{ethreward}. The proposer reward is 1/7 of the total reward for the votes and sync committee's output included by the proposer.

\vspace{-3pt}
\subsection{New Arbitrage Enhancing Chances}
\vspace{-3pt}
The shift of consensus in Ethereum 2.0 has presented new opportunities for arbitrage. The block production process in Ethereum 2.0 has evolved from competition among miners to leader-based block production. A validator is randomly selected for each slot to produce a block as the block proposer (leader). The proposer has absolute power over the construction of the block. Since the block production process does not need to compete with other validators, the proposer can focus on searching for MEV opportunities to increase its revenue. 

\begin{figure}[tp]
\centerline{\includegraphics[width=0.45\textwidth]{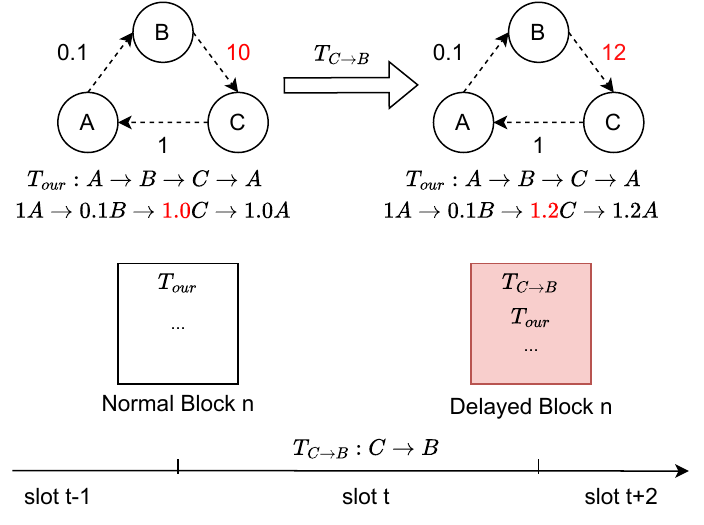}}
\vspace{-3pt}
\caption{Motivating example.}
\label{fig:motivationg_example}
\vspace{-18pt}
\end{figure}

We have observed that if a proposer delays the production of the block, it can observe more transactions, which results in more potential arbitrage opportunities.
The proposer can not only expand the profitability of existing arbitrage opportunities but also build new arbitrage opportunities by ordering subsequent coming transactions. For example, assume that there are three liquidity pools $A$ $\leftrightarrows$ $B$, $B$ $\leftrightarrows$ $C$ and $C$ $\leftrightarrows$ $A$ used for swapping assets $A$, $B$, and $C$ (see Figure~\ref{fig:motivationg_example}). Before slot $t$, transaction $T_{our}$, which swaps tokens along the path $A \rightarrow B \rightarrow C \rightarrow A$, has no arbitrage opportunity.
Following the path, a unit of asset $A$ can be swapped for 0.1\,$B$, then for 1\,$C$, and finally for 1\,A. 
If a transaction, which increases the selling price of any assets of $A$, $B$ and $C$, comes  during slot $t$, then the proposer of slot $t$ can make more profit by delaying block production. For example, a user initiates a transaction $T_{C \rightarrow B}$ to swap asset $C$ for $B$ during slot $t$, which increases the amount of $C$ that a unit of $B$ can swap from 10 to 12.
If the proposer of slot $t$ produces the block normally, it can not profit by only including the transaction $T_{our}$. 
If the proposer of slot $t$ delays block production to slot $t+1$, it can profit 0.2\,$A$ by including user's transaction $T_{C \rightarrow B}$ first, followed by its transaction $T_{our}$.

With the help of ex ante reorganization attacks~\cite{schwarz2022three, neuder2021low} (i.e., pro), blocks can be released with a time delay.
However, Ethereum 2.0 prevents such delay attacks by introducing the proposer boosting mechanism~\cite{proposerboosting}. Suppose the slot’s committee validators receive a newly produced block in the first phase (first 4 seconds) of the current slot. In that case, the block will have a temporary weight of $1/4$ of the total voting weight of the current slot when the validators determine the chain head with the LMD GHOST rule. This requires an adversary controlling at least 1/4 of the voting power 
to launch such an attack.

Adversaries can also delay block production by launching Denial of Service (DoS) attacks on the proposers of other slots so that no new blocks are produced subsequently~\cite{auditEthereum}. 
However, Ethereum's mechanisms such as sentry nodes \cite{sentry}, and SSLE \cite{ssle2022} can prevent such DoS attacks.
The balancing attacks~\cite{neu2022two, neu2021ebb}, which maintain two chains with the same total voting weight so that no transaction can be finalized, may be used to increase the adversary's revenue. 
However, the attack strategies require excellent network control and even violate the Ethereum slashing rule, which results in the adversary losing a significant amount of ETH that has been staked.

In conclusion, existing efforts are unable to achieve the above delayed arbitrage with a limited number of malicious validators under Ethereum 2.0's countermeasures. 
Based on the observation that rational validators are profit-driven, our work proposes the BriDe Arbitrager, which creates a larger time window for proposers to explore arbitrage opportunities by bribing rational validators or proposer of next slot.

\vspace{-3pt}
\subsection{Bribery Attack}
\vspace{-3pt}
Bribery Attack is justifiable in blockchain because the blockchain participants (e.g., miners in PoW and validators in PoS) are rational and profit-driven~\cite{mirkin2020bdos}.
To guarantee blockchain security, blockchains incentivize participants to follow the protocol through various rewards, e.g. block production rewards and transaction fees. However, adversaries can also bribe rational participants to take control of the blockchain within a short period.
Bonneau~\cite{bonneau2016buy} reveals 
that rational participants will accept bribes to maximize their revenue.
After that, several works~\cite{liao2017incentivizing, mccorry2019smart, sun2020model} extend the discussion of bribery attacks to the PoW consensus and implement bribery attacks through whale transaction~\cite{liao2017incentivizing} or smart contracts~\cite{mccorry2019smart}. 


Bribery attack is a potential threat to blockchain security. 
Especially with the popularity of DeFi in Ethereum 2.0, attackers will have more incentives to launch bribery attacks because they can subsidize the cost of bribery attacks through MEV extraction. 
To the best of our knowledge, bribery attacks in Ethereum 2.0 have not to be deeply studied. 
Some works~\cite{david2018ouroboros, ssle2022} mitigate bribery attacks under PoS consensus. 
Ouroboros Praos~\cite{david2018ouroboros} improves the block producer selection method, where the block producer is not published until the block is successfully produced. 
The forthcoming SSLE~\cite{ssle2022} mechanism in Ethereum 2.0 expects to achieve the same effect. However, there is no work to proposing a feasible bribery attack in Ethereum 2.0.

Mechanisms such as SSLE~\cite{ssle2022}, PBS~\cite{pbs2021}, and slashing rule~\cite{buterin2020combining} in Ethereum 2.0 raise new challenges for bribery attacks. 
In this work, we design and implement a bribery-enabled tool called BriDe Arbitrager to increase arbitrage profits, which is still feasible with Ethereum 2.0 countermeasures. We analyse the feasibility of a malicious proposer bribing other validators or proposers separately and derive a bribery strategy, which allows proposers with limited voting power to delay block production. 
We develop a bribery client to automate the whole process and guarantee bribery fairness with a bribery smart contract.
\vspace{-3pt}
\section{Modeling}
\vspace{-3pt}
\subsection{System Model}
\vspace{-3pt}
\label{subsec:system_model}
Assume that there are $N$ validators (i.e., blockchain nodes) $\mathrm{V}_{1},\dotsc,\mathrm{V}_{n},\dotsc,\mathrm{V}_{N}$ in Ethereum 2.0.
Time is divided into epochs.
Each epoch contains $S$ slots, where $S=32$. Each slot lasts for 12 seconds. 
At the beginning of each epoch, the system randomly and uniformly divides validators into 32 consensus committees. Each consensus committee is responsible for consensus in one slot. The block proposer for each slot within this epoch is randomly determined based on the amount of Ether (ETH) staked by the validators.
The block proposer of slot $t$ is denoted as $\mathrm{P}_{t}$.
The honest proposer will publish the block at the beginning of the slot. 
For each slot $t$, validators in the consensus committee $\mathbf{A}_{t} =\{\mathrm{V}_{1},\dotsc,\mathrm{V}_{m},\dotsc,\mathrm{V}_{M}\}$ are responsible for voting for the block, where $M = N/S$ is the number of validator in each consensus committee. 
Each validator makes one vote (attestation) in each epoch. Validators are required to submit their votes $v_{n}$ at the 4th second of the slot for which they are responsible, where the LMD GHOST algorithm is used to identify the block that has the most significant weight of votes in its history as a canonical block. The weight of the vote is positively correlated with the validator's effective balance (i.e., the number of ETH staked). Without loss of generality, we assume that the amount of ETH $d$ staked by each validator is the same. (i.e., $d=32$).

\vspace{3pt}
\noindent
\textbf{Validator Reward.}
In Ethereum 2.0, committee validators are rewarded for: voting for chain head block (for LMD-GHOST), source checkpoint and target checkpoint (for Casper FFG) based on their view of the chain.
However, for validators, our bribery attack only affects the voting rewards for chain head block. 
Therefore, the voting reward of each validator for chain head block is denoted as $R_A = 7/32 \cdot d\cdot r$~\cite{ethreward}, where
$r$ is the base reward (in GWei) per staked Ether~\cite{ethreward}:



%
%

\vspace{-9pt}
\begin{equation}
\begin{aligned}
& r = \frac{10^{9} \times p}{\sqrt{I_{total}}},
\end{aligned}
\vspace{-3pt}
\label{basereward}
\nonumber
\end{equation}
where $I_{total}$ is the total number of ETH (in GWei) of all validators and $p$ is a predefined value, i.e., 64. 
When the validator's vote is correct and is included in the block in time (i.e., it is included in the next block), it receives the corresponding reward.

\vspace{3pt}
\noindent
\textbf{Proposer Reward.}
In Ethereum 2.0, proposers earn the block reward $R_{P}$ by including the votes of other validators (including votes for the head block, source checkpoint, and target checkpoint) and the outputs of the sync committee (who is responsible for signing blocks supporting light clients) in the blocks they produce.
The proposer reward $R_{P}$ is 1/7 of the sum of the total reward for the votes of other validators $R_{A_t}$ and the total reward for the output of the sync committee $R_{Y_t}$ included in the block~\cite{ethreward}. 

\vspace{-3pt}
\begin{equation}
\begin{aligned}
R_{P} = \frac{1}{7}(R_{A_t} + R_{Y_t}) = \frac{1}{7}(\frac{NR_A^{total}}{S} + \frac{Nr}{32}), 
\end{aligned}
\vspace{-3pt}
\label{Rp}
\nonumber
\end{equation}
where $R_A^{total}$ is the total reward of each validator (including voting for source checkpoint, target checkpoint and chain head block~\cite{ethreward}), which can be expressed as $27/32 \cdot d \cdot r$. 

\vspace{3pt}
\noindent
\textbf{DeFi Model.} The system also contains DeFi platforms (i.e., AMMs) that support a set of actions.
For example, the action $o = c_1 \rightarrow c_2$ represents swapping cryptocurrency assets $c_1$ for $c_2$. The corresponding parameter (e.g., the amount of assets swapped) needs to be determined for each action. 
The set of actions supported by all DeFi platforms is denoted $\mathbf{O}$ ($o \in \mathbf{O}$). The set of cryptocurrency assets available to the user for trading is denoted $\mathbf{C}$ ($c \in \mathbf{C}$). 
A sequence of actions executed sequentially forms a path. For example,  $[c_1 \rightarrow c_2 \rightarrow{c_{3}}]$ represents swapping asset $c_1$ for $c_2$ and then swapping asset $c_2$ for $c_3$ on the DeFi platform, which can be executed atomically by a single transaction. 

\vspace{3pt}
\noindent
\textbf{System States.}
The state $s$ of our system consists of the adversaries state, DeFi state, and mempool state. The adversary state is the adversaries' cryptocurrency asset balance profile. DeFi state is the set of DeFi smart contract state variables, i.e., the number of assets in each liquidity pool, which can be read or written by the adversaries. The mempool state is the set of transactions in the mempool pending execution.


\vspace{-3pt}
\subsection{Threat Model}
\vspace{-3pt}
\label{threat model}
We consider an adversary, denoted $\mathbb{T}$, controlling $\alpha_{A}<1/4$ 
proportion of computationally bounded validators.
Validators can execute transactions on a series of DeFi platforms as normal traders and deploy smart contracts on the blockchain. 
When an adversary-controlled validator is selected to be a proposer, it will act like a \emph{malicious proposer}, who may delay block production to the later slot to gain more time for discovering arbitrage opportunities. 
Proposers of the later slots may be controlled by another adversary, denoted $\mathbb{B}$, controlling $\alpha_{B}<1/4$ proportion of validators.
Adversaries aim to search for potential arbitrage opportunities to maximize their revenue (i.e., the balance of their base cryptocurrency assets such as ETH) given the system's state $s$.
Adversaries can observe unconfirmed DeFi transactions in the blockchain network. 
Adversaries can use all available liquidity in the flash loan pool to trade cryptocurrency assets on the DeFi platform~\cite{qin2021attacking}.

\emph{Malicious validators} 
can collude with their malicious proposer. 
They always vote for the block produced by their malicious proposer and can delay voting for blocks and publishing votes as needed. 
Besides malicious validators, we consider the remaining $1-\alpha_{A}-\alpha_{B}$ proportion of validators are \emph{rational validators} who choose to vote for the block that maximizes their revenue based on their view. 
To collect enough votes, adversaries will try to bribe the rational validators to vote for the blocks they produce or to bribe other proposers to delay the block production so that the blocks produced by other proposers cannot collect votes properly. Adversaries may compete to bribe the rational validators of a given slot to obtain arbitrage opportunities for a given period.

We assume that the whole network will receive a message within 4 seconds after it is broadcast. 
This is reasonable based on the evaluation report of Gossipsub~\cite{gossipsub2020}, which is the messaging broadcast protocol for Ethereum 2.0. Moreover, there is a well-connected private network among the malicious validators, where the messaging delay is negligible.

\vspace{-5pt}
\section{Bribery Attack Formalization}
\vspace{-3pt}
\label{sec:Bribery Attack}

In this section, we first formalize the bribery attack and analyse the delay time (when), bribery object (who) and bribery fee (how much) to successfully delay block production.
Based on the analysis, we design an adaptive bribery strategy that allows the adversary who controls a limited fraction of validators to collect enough votes for its delayed block.
Finally, we show that our bribery attack is highly feasible in real cases. 

\vspace{-3pt}
\subsection{Theoretical Analysis}
\vspace{-3pt}
\label{sec:Theoretical Analysis}
In this work, we consider the single-slot bribery attack, i.e., the adversary bribes the validators or proposer of only one slot. 
We leave the detailed analysis of multi-slot bribery attack for future work (as discussed in Section~\ref{sec:limitation}). Intuitively, a multi-slot bribery attack requires a higher cost but does not always lead to more arbitrage profit (or MEV). 

We first discuss the delay time, i.e., when to bribe.
A malicious proposer (adversary) $\mathrm{P}_{t}$ of slot $t$ can delay the block production to slot $t+k$ by bribing validators of slot $t+k$ or the proposer $\mathrm{P}_{t+k}$ of slot $t+k$. However, considering other adversaries, the more slots of delayed block production, the higher the probability of competing with other adversaries (i.e., the proposers of slot $t+1$ to $t+k$ are controlled by other adversaries) for the arbitrage profit $\rho$ between slot $t$ and $t+k$. Intuitively, when multiple adversaries compete against each other, all adversaries eventually spend all their revenues (i.e., block rewards $R_{P}$ and arbitrage profits $\rho$) on bribing other rational validators or proposers. Otherwise, they lose all their revenue if they lose the competition.
As a result, the more slots of delayed block production, the higher the probability of having no revenue.
Therefore, it is \emph{recommended for the malicious proposer $\mathrm{P}_{t}$ to delay block production to the next slot}, which avoids competing with other adversaries.

To decide \emph{who to bribe with how much bribery fee}, we separately analyse the bribery feasibility and requirements for the malicious proposer $\mathrm{P}_{t}$ in terms of bribing the validators and bribing the proposer of slot $t+1$. 
Without loss of generality, we consider the existence of two forks. 
We denote the fork in which the malicious proposer's block is located as the bribery chain $\mathbb{C}_{b}$ and the other fork as the main chain $\mathbb{C}_{m}$~\cite{mirkin2020bdos}. We assume that the malicious proposer $\mathrm{P}_{t}$ is controlled by adversary $\mathbb{T}$ and the proposer $\mathrm{P}_{t+1}$ may be controlled by adversary $\mathbb{B}$.

\begin{figure*}[tp]
  \centering
  \begin{minipage}[t]{0.49\textwidth}
    \centering
    \includegraphics[width=\textwidth]{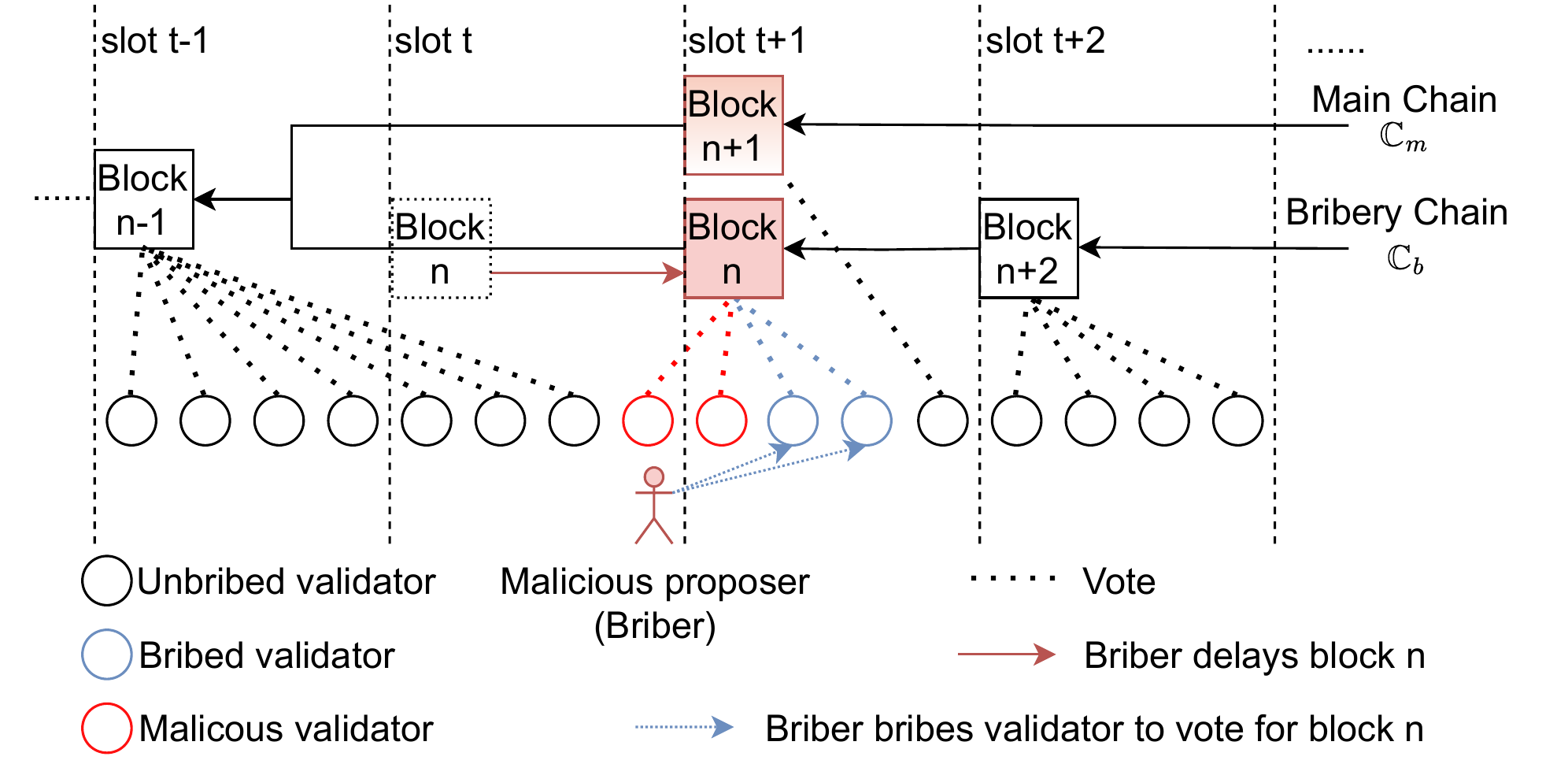}
    \vspace{-15pt}
    \caption{Proposer $\mathrm{P}_{t}$ bribes validators.}
    \label{fig:bribery_attester1}
    
  \end{minipage}
  \begin{minipage}[t]{0.49\textwidth}
    \centering
    \includegraphics[width=\textwidth]{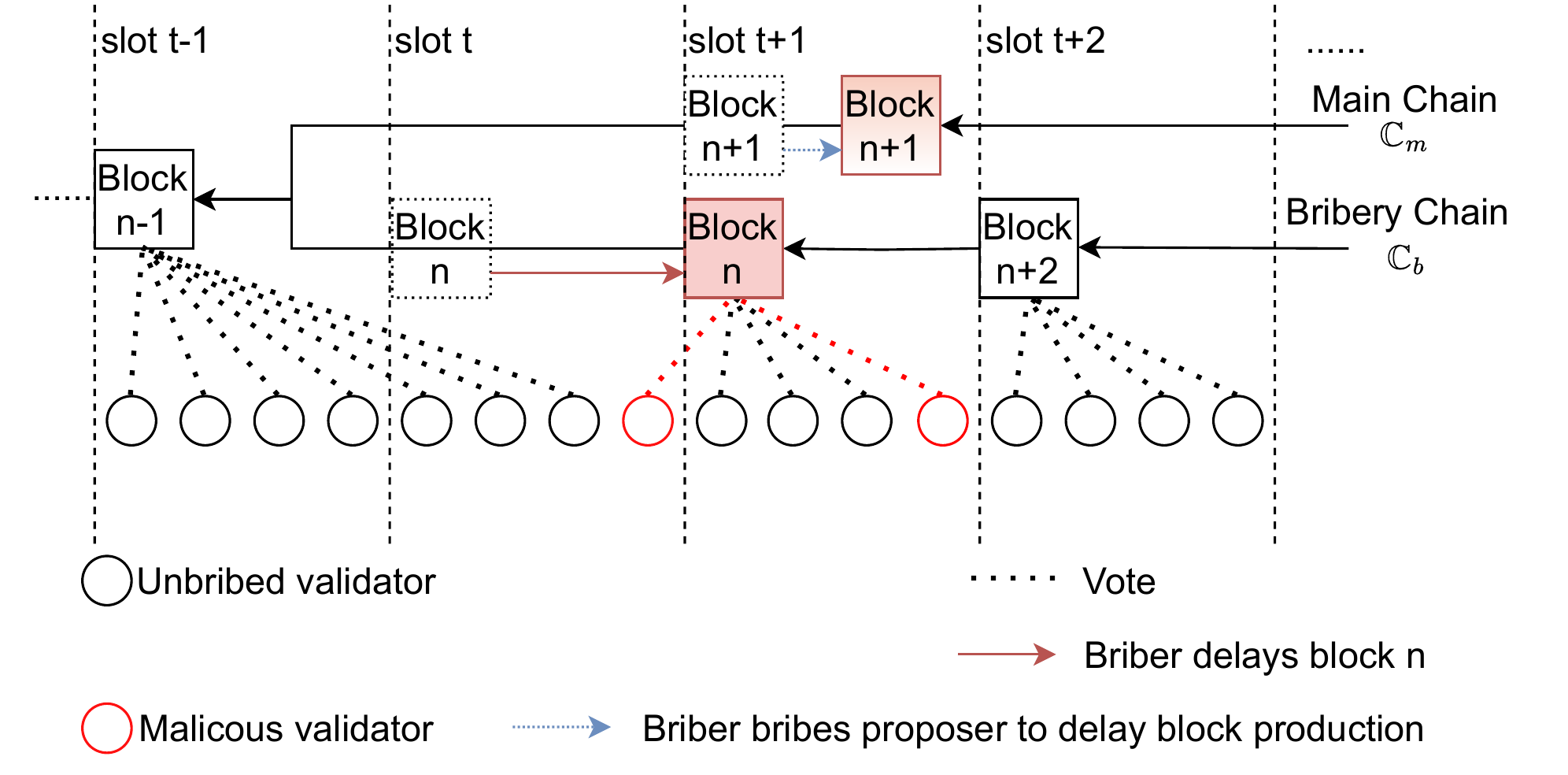}
    \vspace{-15pt}
    \caption{Proposer $\mathrm{P}_{t}$ bribes a proposer.}
    \label{fig:bribery_proposer1}
  \end{minipage}
  \vspace{-15pt}
\end{figure*}


\subsubsection{Bribery Fee of Bribing Validators}
\label{section:Bribe Validators fee}
We now analyse the required \emph{bribery fee} to bribe validators under two adversaries $\mathbb{T}$ and $\mathbb{B}$. 
We model the decision process of the rational validators as a non-cooperative game among them.
Considering the scenarios: the malicious proposer $\mathrm{P}_{t}$ controlled by $\mathbb{T}$ (as briber) delays block production to slot $t+1$ and bribes validators (as bribee) of slot $t+1$ to vote for block $n$ (see Figure~\ref{fig:bribery_attester1}).
The rational validator $\mathrm{V_{q}}$ of slot $t+1$ makes the decision (i.e., votes for which chain), to maximize its utility, based on
the expected vote weights collected by the two chains $\mathbb{C}_{b}$ and $\mathbb{C}_{m}$ after slot $t+1$ in its view, denoted $W_{q}^{b}$ and $W_{q}^{m}$, respectively. 

Bribing validators of slot $t+1$ requires a different bribery fee when the proposer $\mathrm{P}_{t+1}$ is controlled by adversary $\mathbb{B}$ and when it is not.
We first present the utility $U_{q}^{m}$ of the validator $\mathrm{V}_{q}$ voting for the main chain $\mathbb{C}_{m}$ (i.e., not accepting a bribe) and the utility $U_{q}^{b}$ of the validator $\mathrm{V}_{q}$ voting for bribe chain $\mathbb{C}_{b}$ (i.e., accepting a bribe), when the $\mathrm{P}_{t+1}$ is controlled by $\mathbb{B}$. Let the utility when bribed is greater than the utility when not bribed, i.e., $U_{q}^{b}>U_{q}^{m}$, we can derive the bribery fee $\epsilon_{v}$ required by the bribed validator $\mathrm{V}_{q}$ when the $\mathrm{P}_{t+1}$ is controlled by $\mathbb{B}$. 

\noindent
\textbf{Voting Weight. }
The votes $W_{q}^{b}$ that the bribery chain $\mathbb{C}_{b}$ collects after slot $t+1$ consists of the historical votes collected by the $\mathbb{C}_{b}$, denoted $W_{q}^{a}$ (which can be observed by the validator $\mathrm{V_{q}}$), the votes of the malicious validators of slot $t+1$ controlled by $\mathbb{T}$, 
and the votes from other rational validators of slot $t+1$ who vote for the bribery chain $\mathbb{C}_{b}$:

\vspace{-8pt}

\begin{equation}
W_{q}^{b} = W_{q}^{a} + \alpha_{A}\frac{N}{S}+\beta_{q}(1-\alpha_{A}-\alpha_{B})\frac{N}{S},
\label{eqvb}
\vspace{-3pt}
\end{equation}
where $\beta_{q}$ is the proportions of rational validators 
of slot $t+1$ who vote for $\mathbb{C}_{b}$ predicted by the validator $\mathrm{V}_{q}$. 
Recall that 
$N/S$ is the number of validators in each slot. 
The rest are rational validators except for the $\alpha_{A}$ of validators controlled by $\mathbb{T}$, and the $\alpha_{B}$ of validators controlled by $\mathbb{B}$.

Correspondingly, all votes that the main chain $\mathbb{C}_{m}$ collects after slot $t+1$, i.e., $W_{q}^{m}$, consists of the historical votes already collected by $\mathbb{C}_{m}$ (denoted $W_{q}^{v}$), the votes of the malicious validators of slot $t+1$ controlled by $\mathbb{B}$, and the votes from other validators of slot $t+1$ who vote for the main chain $\mathbb{C}_{m}$: 

\vspace{-10pt}
\begin{equation}
W_{q}^{m} = W_{q}^{v} + \alpha_{B}\frac{N}{S}+(1-\beta_{q})(1-\alpha_{A}-\alpha_{B})\frac{N}{S}.
\label{eqvm}
\end{equation}

\noindent
\textbf{Validator Utility. }
To compete for arbitrage profit $\rho$ from slot $t$ to slot $t+1$, both $\mathbb{T}$ and $\mathbb{B}$ try to bribe the rational validators of slot $t+1$. 
When $\mathrm{V}_{q}$ votes for $\mathbb{C}_{m}$, the total vote weights of $\mathbb{C}_{m}$ is ${W'}_{q}^{m} = W_{q}^{m} + 1$. At this point, if 
$\mathbb{C}_{m}$ becomes the canonical chain, $\mathrm{V}_{q}$ can obtain the voting reward $R_{A}$ and possible bribery fee $\epsilon_{m}$ provided by $\mathbb{B}$. Therefore, $\mathrm{V}_{q}$'s utility of voting for $\mathbb{C}_{m}$, denoted $U_{q}^{m}$, is 
\vspace{-3pt}
\begin{equation}
\begin{aligned}
& U_{q}^{m} = Pr\{{W'}_{q}^{m}-W_{q}^{b}>0\}(R_{A}+\epsilon_{m})\\
& = Pr\{\beta_{q}<\frac{S(W_{q}^{v}-W_{q}^{a}+1)+N(1-2\alpha_{A})}{2N(1-\alpha_{A}-\alpha_{B})}\}(R_{A}+\epsilon_{m}).
\end{aligned}
\label{equvm1}
\end{equation}
$Pr\{{W'}_{q}^{m}-W_{q}^{b}>0\}$ is the probability that the vote weights of $\mathbb{C}_{m}$ are larger than the the vote weights of $\mathbb{C}_{b}$, which depends on $\beta_{q}$.
Because rational validators do not know each other's information, we treat $\beta_{q}$ as a random variable uniformly distributed between [0, 1]. Therefore, we have

\vspace{-3pt}
\begin{equation}
\begin{aligned}
 Pr\{\beta_{q}<&\frac{S(W_{q}^{v}-W_{q}^{a}+1)+N(1-2\alpha_{A})}{2N(1-\alpha_{A}-\alpha_{B})}\} =\\
& \frac{S(W_{q}^{v}-W_{q}^{a}+1)+N(1-2\alpha_{A})}{2N(1-\alpha_{A}-\alpha_{B})}.
\end{aligned}
\label{eq:probability of cm}
\end{equation}
\vspace{-3pt}
Combine formulation~\eqref{equvm1} with~\eqref{eq:probability of cm}, we can obtain
\begin{equation}
\begin{aligned}
& U_{q}^{m} = \frac{S(W_{q}^{v}-W_{q}^{a}+1)+N(1-2\alpha_{A})}{2N(1-\alpha_{A}-\alpha_{B})}(R_{A}+\epsilon_{m}).
\end{aligned}
\label{equvm}
\end{equation}
\vspace{-3pt}

Similarly, when $\mathrm{V}_{q}$ votes for $\mathbb{C}_{b}$, the total vote weights of $\mathbb{C}_{b}$ is ${W'}_{q}^{b} = W_{q}^{b} + 1$. If the $\mathbb{C}_{b}$ becomes the canonical chain, $\mathrm{V}_{q}$ 
receive voting reward $R_{A}$, 
and bribery fee $\epsilon_{v}$ offered by the briber $\mathrm{P}_{t}$ (i.e., the adversary $\mathbb{T}$). The expected utility of $\mathrm{V}_{q}$ for voting for $\mathbb{C}_{b}$, denoted $U_{q}^{b}$, can be expressed as:
\vspace{-3pt}
\begin{equation}
\begin{aligned}
& U_{q}^{b} = Pr\{{W'}_{q}^{b}-W_{q}^{m}>0\}(R_{A}+\epsilon_{v})\\
& = Pr\{\beta_{q}>\frac{S(W_{q}^{v}-W_{q}^{a}-1)+N(1-2\alpha_{A})}{2N(1-\alpha_{A}-\alpha_{B})}\}(R_{A}+\epsilon_{v})\\
& = \frac{S(1-W_{q}^{v}+W_{q}^{a})+N(1-2\alpha_{B})}{2N(1-\alpha_{A}-\alpha_{B})}(R_{A}+\epsilon_{v}).
\end{aligned}
\label{equvb}
\end{equation}

\noindent
\textbf{Bribery Fee Requirement. }
When the $\mathrm{P}_{t+1}$ is controlled by $\mathbb{B}$, if $U_{q}^{b} > U_{q}^{m}$, i.e., bribery fee $\epsilon_{v}$ paid by the proposer $\mathrm{P}_{t}$  to each validator satisfies the minimal bribery fee requirement (Equation ~\eqref{bribery_fee}), $\mathrm{V}_{q}$ choose to accept the bribe (vote for $\mathbb{C}_{b}$). $\mathrm{P}_{t}$ thus successfully delays block production.
\vspace{-3pt}
\begin{subequations}
\begin{align}
\epsilon_{v} &> \frac{2SR_{A}(W_{q}^{v}-W_{q}^{a})+2NR_{A}(\alpha_{B}-\alpha_{A})}{S(1-W_{q}^{v}+W_{q}^{a})+N(1-2\alpha_{B})} \nonumber \\
            & +\frac{\epsilon_{m}(S(1+W_{q}^{v}-W_{q}^{a})+N(1-2\alpha_{A}))}{S(1-W_{q}^{v}+W_{q}^{a})+N(1-2\alpha_{B})}, \\
\epsilon_{v} &> 0.
\label{bribery_fee}
\end{align}
\end{subequations}
\vspace{-12pt}

Be aware that the bribery fee is not negative.

To derive the bribery fee $\epsilon_{v}'$ required for the validator when the proposer $\mathrm{P}_{t+1}$ is not controlled by $\mathbb{B}$, we take $\alpha_{B}$ and $\epsilon_{m}$ to be 0.
In this case, the validators controlled by $\mathbb{B}$ will act as rational validators.
Substituting $\alpha_{B} = 0$ and $\epsilon_{m} = 0$ into Equation~\eqref{bribery_fee}, 
We obtain that the rational validators of slot $t+1$ accepts the bribe as long as bribery fee $\epsilon_{v}'$ satisfies Equation~\eqref{bribery_fee_normal}, which maintains $U_{q}^{b} > U_{q}^{m}$.

\vspace{-8pt}
\begin{subequations}
\begin{align}
\epsilon_{v}' &> \frac{2SR_{A}(W_{q}^{v}-W_{q}^{a})-2NR_{A}\alpha_{A}}{S(1-W_{q}^{v}+W_{q}^{a})+N}, \\
\epsilon_{v}' &> 0.
\end{align}
\label{bribery_fee_normal}
\end{subequations}
\vspace{-5pt}


\subsubsection{Bribery Fee of Bribing Proposers}
\vspace{-3pt}
\label{section: Bribe Proposers}
Bribing proposer of slot $t+1$ also requires a different bribery fee when the proposer $\mathrm{P}_{t+1}$ is controlled by adversary $\mathbb{B}$ and when it is not. 
Similar to Section~\ref{section:Bribe Validators fee}, 
we now analyse the required \emph{bribery fee} of bribing the proposer $\mathrm{P}_{t+1}$ when it is not controlled by $\mathbb{B}$. 

Considering the scenarios: the malicious proposer $\mathrm{P}_{t}$ controlled by the adversary $\mathbb{T}$ delays block production to slot $t+1$ and bribes the rational proposer $\mathrm{P}_{t+1}$ to give up the block proposing opportunity, i.e., produce the block after the 4th second of slot $t+1$ (see Figure~\ref{fig:bribery_proposer1}).
In this case, if proposer $\mathrm{P}_{t+1}$ accepts the bribe to produce block after the 4th second of slot $t+1$, proposer $\mathrm{P}_{t+1}$'s block would not trigger the proposer boosting mechanism~\cite{proposerboosting}. 
Since the validators of slot $t+1$ do not receive block $n+1$ before the 4th second of slot $t+1$, but do receive block $n$, they vote for block $n$. Therefore, block $n$ collects enough votes.

\noindent
\textbf{Proposer Utility. } To maximize its utility, there are two possible choices for the rational proposer $\mathrm{P}_{t+1}$: produce block honestly, or produce blocks after the 4th second of slot $t+1$ (i.e., accept the bribe). 
When $\mathrm{P}_{t+1}$ produces block honestly, its expected utility $U_{\mathrm{P}_{t+1}}^{m}$ is the product of the probability $Pr\{\text{finalizing}\ \mathbb{C}_{m}\}$ that the main chain $\mathbb{C}_{m}$ eventually becomes the canonical chain and the expected block reward $R_{P}$:

\vspace{-10pt}

\begin{equation}
U_{\mathrm{P}_{t+1}}^{m} = Pr\{\text{finalizing}\ \mathbb{C}_{m}\}R_{P}.
\label{equpm}
\end{equation}

\vspace{-3pt}

When accepting the bribe, proposer $\mathrm{P}_{t+1}$ receives the bribery fee $\epsilon_{p}'$ offered by the briber $\mathrm{P}_{t}$ if $\mathbb{C}_{b}$ becomes canonical chain. Otherwise, proposer $\mathrm{P}_{t+1}$ receives the block reward $R_{P}$. Therefore, its expected utility $U_{\mathrm{P}_{t+1}}^{b}$ for $\mathrm{P}_{t+1}$ can be expressed as:

\vspace{-15pt}

\begin{equation}
U_{\mathrm{P}_{t+1}}^{b} = Pr\{\text{finalizing}\ \mathbb{C}_{b}\}\epsilon_{p}' + (1-Pr\{\text{finalizing}\ \mathbb{C}_{b}\})R_{P}.
\label{equpb}
\end{equation}
Similar to the Equation~\eqref{equpm}, $Pr\{\text{finalizing}\ \mathbb{C}_{b}\}$ is the probability that $\mathbb{C}_{b}$ becomes canonical chain. 

\noindent
\textbf{Bribery Fee Requirement.} Obviously, when the bribery fee $\epsilon_{p}' > R_{P}$, we always guarantee that $U_{\mathrm{P}_{t+1}}^{b} > U_{\mathrm{P}_{t+1}}^{m}$, i.e., accepting a bribe yields a higher utility. 
Therefore, the proposer $\mathrm{P}_{t+1}$ will accept the bribe.
The malicious proposer (adversary $\mathbb{T}$) is required to pay a bribery fee of at least $R_{P}$ when $\mathrm{P}_{t+1}$ is not controlled by $\mathbb{B}$.
However, when the proposer $\mathrm{P}_{t+1}$ is controlled by the adversary $\mathbb{B}$ who tries to extract arbitrage profits $\rho$, the malicious proposer is required to pay a bribery fee $\epsilon_{p}$ that covers both the block reward $R_{P}$ and the arbitrage profits $\rho$, i.e., $\epsilon_{p} > R_{P} + \rho$, to successfully delay block production. 

\vspace{-3pt}
\subsection{Bribery Strategy Selection}
\vspace{-3pt}
\label{section: optimal Bribe}
We now derive the bribery strategy that enables the adversary $\mathbb{T}$ to delay block production. The adversary $\mathbb{T}$ select the bribery object with the lowest total bribery cost based on the number of malicious validators in the control of the adversary 
$\mathbb{T}$ and the number of validators that can be bribed.

We first define the total bribery cost $\tau$ of the adversary $\mathbb{T}$.
As will be elaborated in Section \ref{subsec:briberySC}, the bribees automatically withdraw the bribery fee through our designed bribery smart contract.
This, however, requires the bribees to pay transaction fees. 
To motivate rational proposers and validators to accept bribes, the malicious proposer (briber) needs to subsidise each bribee with a corresponding withdrawal fee $\theta$. 
Therefore, the total bribery cost $\tau$ for the briber consists of the total bribery fee $\epsilon^{total}$ and the total withdrawal fee $\theta^{total}$, i.e., $\tau=\epsilon^{total}+\theta^{total}$, where $\theta^{total}$ is the product of the number of bribees and $\theta$.

\subsubsection{Bribery Cost of Bribing Validators}
\label{subsec:Cost of Bribing Validators}
We now derive the bribery cost of bribing the validators under two adversaries $\mathbb{T}$ and $\mathbb{B}$. 
The bribery cost of bribing the validators is different when the proposer $\mathrm{P}_{t+1}$ is controlled by $\mathbb{B}$ and when it is not. 
The malicious proposer $\mathrm{P}_{t}$ needs to bribe $\beta$ proportion of rational validators of slot $t+1$ to successfully launch the bribery attack. 
If the proposer of slot $t+1$ is not controlled by $\mathbb{B}$, $\mathrm{P}_{t}$ needs to pay a bribery fee of $\epsilon_{v}'$ to each bribed validator. 
Thus, $\mathrm{P}_{t}$ has to pay the total bribery fee
\vspace{-3pt}
\begin{equation}
\begin{aligned}
{\epsilon'}_{v}^{total} = \beta \cdot (1-\alpha_{A}-\alpha_{B}) \cdot \frac{N}{S} \cdot \epsilon_{v}'.
\end{aligned}
\label{bribery_totalfee}
\end{equation}
Both $\beta$ and $\epsilon_{v}'$ are related to the historical voting weights of the main chain $\mathbb{C}_{m}$ (denoted $W^{v}$) and the historical voting weights of the bribery chain $\mathbb{C}_{b}$ (denoted $W^{a}$). 

In the scenario mentioned in Section~\ref{section:Bribe Validators fee}, the historical votes collected by $\mathbb{C}_{m}$ are 0, i.e, the block $n+1$ produced by proposer $\mathrm{P}_{t+1}$ does not receive any vote at the beginning of slot $t+1$.
The historical votes collected by $\mathbb{C}_{b}$ are the votes of all malicious validators of slot $t$. We thus have $W^{v}=0$ and $W^{a}=\alpha_{A}\frac{N}{S}$. 
Assuming that the validator view is the same as the global view, i.e., $W^{b}=W_{q}^{b}$, $W^{m}=W_{q}^{m}$, $W^{v}=W_{q}^{v}$ and $W^{a}=W_{q}^{a}$. 
Therefore, we derive the following theorem:
\begin{theorem}
For the malicious proposer $\mathrm{P}_{t}$, successfully delaying block production, i.e., the bribery chain $\mathbb{C}_{b}$ collects a larger voting weight than that collected by the main chain $\mathbb{C}_{m}$, requires that the proportions of bribed rational validators $\beta$ satisfies:

\vspace{-5pt}
\begin{equation}
\begin{aligned}
\beta>\frac{N-3N\alpha_{A}}{2N(1-\alpha_{A}-\alpha_{B})}.
\end{aligned}
\label{bribery_num1}
\end{equation}
\vspace{-10pt}
\label{th:bribery num}
\end{theorem}
\vspace{-15pt}
\begin{proof}
Let the vote weights of the bribery chain $\mathbb{C}_{b}$ be larger than the vote weights of the main chain $\mathbb{C}_{m}$, i.e, $W^{b}>W^{m}$. Then, we have $W^{a}-W^{v}+(\alpha_{A}-\alpha_{B})\frac{N}{S} + (2\beta-1) (1-\alpha_{A}-\alpha_{B})\frac{N}{S}>0$. Solving the above formula, we obtain that $\beta>(N-3N\alpha_{A})/(2N(1-\alpha_{A}-\alpha_{B}))$.
\end{proof}

\vspace{-5pt}
Substituting $W^{v}$ and $W^{a}$ into Equation~\eqref{bribery_fee_normal}, we obtain the bribery fee $\epsilon_{v}'$ when the proposer of slot $t+1$ is not controlled by $\mathbb{B}$ should satisfy:
\vspace{-3pt}
\begin{subequations}
\label{bribery_fee1}
\begin{align}
\epsilon_{v}' &> \frac{-4NR_{A}\alpha_{A}}{S+N(1+\alpha_{A})}, \label{bribery_fee1a}\\
\epsilon_{v}' &> 0. \label{bribery_fee1b}
\end{align}
\vspace{-15pt}
\end{subequations}



On the other hand, there exists $\alpha_{B}$ probability that the proposer of slot $t+1$ is controlled by $\mathbb{B}$. In this case, adversaries $\mathbb{T}$ and $\mathbb{B}$ always compete to bribe the rational validators of slot $t+1$ to vote for the block they produced.
Otherwise, they may lose the block reward $R_{P}$ and the arbitrage profit $\rho$ from slot $t$ to $t+1$. 
In the worst case, $\mathbb{T}$ will use all of its block reward $R_{P}$ and arbitrage profit $\rho$ to bribe the rational validators of slot $t+1$, which results in no revenue for $\mathbb{T}$. 
Therefore, we can derive the expected total bribery fee $\overline{\epsilon_{v}^{total}}$ for $\mathbb{T}$ to bribe validators of slot $t+1$,
\vspace{-3pt}
\begin{equation}
\begin{aligned}
\overline{\epsilon_{v}^{total}} = (1-\alpha_{B}){\epsilon'}_{v}^{total} + \alpha_{B}(R_{P}+\rho).
\end{aligned}
\label{total bribery fee v}
\vspace{-3pt}
\end{equation}
The right side of the Formula~\eqref{bribery_fee1a} is negative. Therefore, the bribery fee $\epsilon_{v}'$ paid to each validator \emph{\textbf{can be any negligible positive value $\sigma>0$}}, so does the total bribery fee ${\epsilon'}_{v}^{total}$.
Considering practical aspects, we suppose that the briber $\mathrm{P}_{t}$ would pay 1\,GWei bribery fee to each bribed validator of slot $t+1$. 
Meanwhile, the malicious proposer $\mathrm{P}_{t}$ need to subsidise total withdrawal fee $\theta_{v}^{total}$ to $\beta$ proportion of rational validators of slot $t+1$,
\begin{equation}
\begin{aligned}
\theta_{v}^{total} = \beta \cdot (1-\alpha_{A}-\alpha_{B}) \cdot \frac{N}{S} \cdot \theta.
\end{aligned}
\label{total withdrawal fee}
\vspace{-8pt}
\end{equation}
\vspace{-5pt}

Therefore, the total bribery cost of bribing validators is the sum of the expected total bribery fee $\overline{\epsilon_{v}^{total}}$ and total withdrawal fee $\theta_{v}^{total}$ of bribing validators of slot $t+1$.
\begin{equation}
\begin{aligned}
\tau_{v} = &\overline{\epsilon_{v}^{total}}  + \theta_{v}^{total} \\
= &\sigma + \alpha_{B}(R_{P}+\rho) + \frac{N-3N\alpha_{A}}{2S} \cdot \theta.
\end{aligned}
\label{bribery cost v}
\end{equation}
\subsubsection{Bribery Cost of Bribing Proposer}
\label{subsec:Cost of Bribing Proposer}
The malicious proposer $\mathrm{P}_{t}$ can bribe the proposer $\mathrm{P}_{t+1}$ to delay block production  to slot $t+1$. $\mathrm{P}_{t}$ is required to pay the bribery fee $\epsilon_{p}'$ at least $R_{P}$ for bribing rational proposer $\mathrm{P}_{t+1}$.
On the other hand, if $\mathrm{P}_{t+1}$ is malicious proposer controlled by $\mathbb{B}$, bribery fee $\epsilon_{p}$ should satisfy that $\epsilon_{p} > R_{P} + \rho$.
 Therefore, we can derive the expected total bribery fee $\overline{\epsilon_{p}^{total}}$ for $\mathbb{T}$ to bribe proposer $\mathrm{P}_{t+1}$:
\vspace{-3pt}
\begin{equation}
\begin{aligned}
\overline{\epsilon_{p}^{total}} &= (1-\alpha_{B})R_{P} + \alpha_{B}(R_{P}+\rho) = R_{P} + \alpha_{B}\rho.
\end{aligned}
\label{total bribery fee p}
\vspace{-3pt}
\end{equation}
Only the proposer $\mathrm{P}_{t+1}$ needs to be bribed and subsidized the withdrawal fee, so $\theta_{p}^{total} = \theta$.
Therefore, the total bribery cost of bribing proposer $\mathrm{P}_{t+1}$ is the sum of the expected total bribery fee $\overline{\epsilon_{p}^{total}}$ and total withdrawal fee $\theta_{p}^{total}$ of bribing proposer of slot $t+1$.
\begin{equation}
\begin{aligned}
\tau_{p} = &\overline{\epsilon_{p}^{total}}  + \theta_{p}^{total} 
= R_{P} + \alpha_{B}\rho + \theta.
\end{aligned}
\label{bribery cost p}
\vspace{-3pt}
\end{equation}


\subsubsection{Adaptive Bribery Strategy}
To derive the adaptive bribery strategy for adversary $\mathbb{T}$, we need to determine whether bribing validators or bribing proposers has a lower bribery cost.
Combining the several scenarios of bribing validators and proposer described above, we derive the following theorem:
\vspace{-5pt}
\begin{theorem}
$\forall {\alpha_{A}, \alpha_{B} \in (0,0.25)}$, the bribery cost of bribing validators is less than the bribery cost of bribing proposer, i.e., $\tau_{v}< \tau_{p}$, when $0<N<18358621$.
\vspace{-4pt}
\label{th:bribery cost}
\end{theorem}
\vspace{-8pt}
\begin{proof}
Let $f(\alpha_{A}, \alpha_{B}) = \tau_{v} - \tau_{p} = \sigma + (\alpha_{B}-1)R_{P} + (N-2S-3N\alpha_{A})/(2S)\cdot \theta$.
Because $\frac{\partial f(\alpha_{A}, \alpha_{B})}{\partial \alpha_{A}} = -3N < 0$ and $\frac{\partial f(\alpha_{A}, \alpha_{B})}{\partial \alpha_{B}} = R_{P} > 0$, $f(\alpha_{A}, \alpha_{B})$ is monotonically decreasing with $\alpha_{A}$ and monotonically increasing with $\alpha_{B}$. Therefore, $f(\alpha_{A}, \alpha_{B})$ is maximum at $\alpha_{A}=0, \alpha_{B}=0.25$ for $\alpha_{A},\alpha_{B} \in (0,0.25)$. 
If $f(0, 0.25)<0$, then we have $f(\alpha_{A}, \alpha_{B}) < 0$ $\forall \alpha_{A}, \alpha_{B} \in (0,0.25)$. $f(0, 0.25)=\sigma -0.75R_{P} + (N-2S)/(2S) \cdot \theta$, where $R_{P}=\sqrt{2N/10^9}$ and $\sigma$ is a negligible positive value. Therefore, solving for $f(\alpha_{A}, \alpha_{B}) < 0$, is equivalent to solving for $f^2(\alpha_{A}, \alpha_{B}) < 0$, i.e., $N^2-\left[4S+(9S^2)/(2 \times 10^9 \times \theta^2)\right] N+4S^2<0$. With the development and tests of the bribery contract, we verify that $\theta$ is expected to be $5.01*10^{-7}$ ETH. We then get $f(0, 0.25)<0$ when $0<N<18358621$. The number $N$ of validators in Ethereum will be in this range for a long time~\cite{beaconcha}. Therefore, $f(\alpha_{A}, \alpha_{B}) < 0$, i.e., $\tau_{v}< \tau_{p}$ holds for all $\alpha_{A},\alpha_{B} \in (0,0.25)$, when $0<N<18358621$.
\end{proof}
\vspace{-8pt}

\textbf{Summary of the Adaptive Bribery Strategy.} Based on Theorem~\ref{th:bribery cost}, we learn that bribing validators has a lower bribery cost. However, based on Theorem~\ref{th:bribery num}, we know that successfully delaying block production requires that the proportion  of bribed rational validators $\beta$ satisfies Equation~\eqref{bribery_num1}. While bribing the proposer incurs a higher bribery cost, it only requires that the proposer of the next slot be bribable, which is less than the number of bribe objects bribing validators requires. Therefore, we derive the adaptive bribery strategy for adversary $\mathbb{T}$: the malicious proposer $\mathrm{P}_{t}$ bribes the validators of slot $t+1$ to vote for $\mathrm{P}_{t}$'s delayed block $n$, when the proportions of bribable rational validators greater than the minimal required proportions (Equation~\eqref{bribery_num1}). Otherwise, when the proposer of slot $t+1$ is bribable, the malicious proposer $\mathrm{P}_{t}$ bribes the proposer of slot $t+1$ to delay block production for 4 seconds. 
As a result, the adversary $\mathbb{T}$ who controls a limited fraction of validators can delay the production of block $n$ to slot $t+1$ in various scenarios. 
Note that when the proportion $\alpha_{A}$ of validators controlled by the adversary $\mathbb{T}$ is greater than 1/4, $\mathbb{T}$ can delay block production via ex ante reorganization attack~\cite{schwarz2022three}, but this is outside the scope of this paper.

\vspace{3pt}
\noindent
\textbf{Free of Slashing.}
The adaptive bribery strategy introduces three kinds of malicious behaviour: malicious proposers and bribed proposers delaying block production, malicious validators delaying the voting process, and bribed validators disregarding the fork choice rule to vote for the delayed block. 
Honest validators will consider the network delay as the cause of these abnormal behaviours.  
Since neither the proposer's delayed block production nor the validators' votes for the delayed block violate the protocol's slashing rules (proposing two conflicting blocks or voting for two conflicting blocks simultaneously), the malicious proposers and the validators are not penalized.

\vspace{-5pt}
\section{Delayed Transaction Ordering Algorithm}
\vspace{-3pt}
\label{sec:Delayed Arbitrage}
The arbitrage profits that a proposer can extract are associated with the arbitrage detection algorithm and transaction flow that it uses.
With the help of the adaptive bribery strategy, a malicious proposer can delay block production for one slot (i.e., 12s)
to observe a longer period of the blockchain state (i.e., prices of assets on DEX and transactions in mempool), which means a larger transaction flow can be collected. 
In this section, we propose the Delayed Transaction Ordering Algorithm (DTOA) that can utilize this extended period and larger transaction flow to discover more profitable arbitrage opportunities.


\vspace{-7pt}
\subsection{DTOA Overview}
\vspace{-3pt}
\label{subsec:DTOA_overview}

The existing arbitrage detection algorithm~\cite{zhou2021just} is designed for high frequency traders, so it greedily extracts the corresponding arbitrage profit whenever they find an arbitrage opportunity for speed and often misses highly profitable arbitrage opportunities. Different from the existing arbitrage algorithm, the DTOA is an arbitrage algorithm designed specifically for proposers. Therefore, the DTOA focuses on finding better arbitrage opportunities rather than finding arbitrage opportunities faster. Moreover, the proposer can determine the order of transitions in the block. Thus, the DTOA should be able to use the ability to order transactions to increase arbitrage profits. 

Arbitrage detection can be viewed as a cycle detection problem for the weighted directed graph.
We introduce the following notations to aid in understanding our algorithm: 

\noindent
\textbf{Nodes:} Each node (vertex) in the directed graph represents a cryptocurrency asset ($c \in \mathbf{C}$). The collection of all nodes is denoted as set $\mathbf{V}$.

\noindent
\textbf{Directed edges:} Each edge $e_{i,j}$ in the directed graph, which points from asset $c_i$ to $c_j$, represents the existence of a DeFi AMM platform that allows traders to purchase  $c_j$ with $c_i$. The collection of all edges is denoted as set $\mathbf{E}$.

\noindent
\textbf{Edge weight:} 
The weight $w_{i,j}$ of the edge $e_{i,j}$ represents the current approximate best price at which a trader can purchase $c_j$ with a sufficiently small (approaching with 0) amount of $c_i$ on all DeFi AMM platforms. It can be expressed as the ratio of the reserve of assets $c_j$ and assets $c_i$ in the liquidity pool that has the best price, i.e., $w_{i,j} = \frac{Q_{c_j}}{Q_{c_i}}$. If no liquidity pool supports swapping $c_i$ and $c_j$, then $w_{i,j}$ is set to 0.

\noindent
\textbf{Arbitrage:} 
If $w_{1,2} \times \dots \times w_{k-1,k} > 1$, then an arbitrage opportunity exists for path $[c_1\xrightarrow{w_{1,2}}c_2 \dots c_{k-1}\xrightarrow{w_{k-1,k}}{c_{k}}]$.

\noindent
\textbf{Arbitrage strategy:} An arbitrage strategy consists of a set of sequential transactions, where the last transaction is our arbitrage transaction, and the parameter (trading amount of initial trading asset, e.g. $c_1$) for our arbitrage transaction. The arbitrage strategy prioritises the ordering (executing) of other users' trades $T_{others}$ (see Section~\ref{Delayed Arbitrage Arbitrage Details}'s Transaction Ordering) before including (executing) arbitrage transaction.

\noindent
\textbf{Arbitrage strategy sequence finding:} Our objective is to find a series of arbitrage strategy sequences to maximize adversary $\mathbb{T}$'s balance of the base cryptocurrency asset. Each arbitrage strategy sequence consists of several sequential arbitrage strategies.



We input the blockchain state $s$, the node set $\mathbf{V}$ and the edge set $\mathbf{E}$ to the DTOA, which outputs an arbitrage strategy sequence.
Specifically, the DTOA first use DFS to find all potential arbitrage paths (cycles).
For each arbitrage path, DTOA prioritises the execution of the specific transactions $T_{others}$ accumulated in the memory pool to increase the selling price of the asset involved in the path. Then, DTOA selects the most profitable arbitrage path and composes the transaction sequence (arbitrage strategy) in the corresponding order, i.e., first $T_{others}$ then the arbitrage transaction. The DTOA repeats the above steps until no new arbitrage opportunities are identified.
On a high level, the DTOA consists of the following five key components.


\noindent
\textbf{(1) Liquidity Pool Selection.} A heuristic liquidity pool selection strategy is used to select liquidity pools that are more probable to have arbitrage opportunities. The DTOA efficiently searches for arbitrage opportunities in a reduced search space (state $s$).

\noindent
\textbf{(2) Cycle Detection.} Build the graph $g$ based on the state $s$. Then detect all cycles in the graph using the DFS algorithm and construct the path set $\mathbf{path}$.

\noindent
\textbf{(3) Transaction Ordering.} Order transactions based on each path ($path \in \mathbf{path}$) separately to optimize the trade price of the assets in the path.

\noindent
\textbf{(4) Trading Amount Optimization and Strategy Determination.} Determine the optimal amount of initial trading asset for each path $path$. Then identify the optimal arbitrage strategy that has the most revenue.


\vspace{2pt}
\noindent
\textbf{(5) Validation and State Updating.}  Execute the arbitrage strategy locally to validate if the arbitrage strategy is profitable. If profitable, include the transaction into the block according to the arbitrage strategy. The algorithm then synchronizes the state $s$ to the state after executing the arbitrage strategy.

\vspace{-5pt}
\subsection{DTOA Designs}
\vspace{-3pt}
\label{Delayed Arbitrage Arbitrage Details}
We now discuss the detailed designs of the DTOA, which tries to discover and enhance potential cyclic arbitrage opportunities. 

\vspace{1pt}
\noindent
\textbf{Liquidity Pool Selection.}
The number of optional liquidity pools is so large that it is impractical to construct a graph with the complete state. 
Therefore, we design some heuristics to guide the selection of liquidity pools.
We aim to \emph{select liquidity pools more likely to have arbitrage opportunities with higher profits.}
We focus on searching for arbitrage opportunities within the selected liquidity pools.
The selected liquidity pools comprise a smaller and more practical system state (search space) than the complete state. 
The DTOA algorithm efficiently searches for more profitable arbitrage opportunities in the reduced search space. 
In addition, we focus on constant product market makers so that we can use our trading amount optimization technique.

\noindent
\textit{\textbf{Heuristic 1:}} Liquidity pools with \emph{high trading volumes} have a greater probability of taking advantage of other users' transactions to enhance arbitrage.

\noindent
\textit{\textbf{Heuristic 2:}} The same size transaction can cause greater price volatility for liquidity pools with \emph{lower liquidity}. 

\vspace{1pt}
\noindent
\textbf{Cycle Detection.}
In order to find the optimal arbitrage opportunity, we propose to use the DFS algorithm in the DTOA to find all cycles in the graph $g$. We then apply transaction ordering for each cycle. Finally, we select the most profitable one in strategy determination. 
However, DFS is not guaranteed to find an arbitrage cycle.
Therefore, for each cycle, we form two paths by forward traversal and backward traversal and finally choose one with arbitrage opportunities. 
The rationale of this solution is based on the following theorem we observed:

\vspace{-5pt}
\begin{theorem}
If an arbitrage opportunity does not exist for $path = [c_1\xrightarrow{w_{1,2}}c_2 \dots c_{k-1}\xrightarrow{w_{k-1,k}}{c_{k}}]$, i.e., $0 < w_{1,2} \times \dots \times w_{k-1,k} < 1$, then an arbitrage opportunity exists for the inverse path $\overline{path} = [c_k\xrightarrow{w_{k,k-1}}c_{k-1} \dots c_{2}\xrightarrow{w_{2,1}}{c_{1}}]$, i.e., $w_{k,k-1} \times \dots \times w_{2,1} > 1$ unless the assets prices are the same on different liquidity pools.
\vspace{-4pt}
\end{theorem}
\vspace{-4pt}
\begin{proof}
We have $w_{i,j} = w_{j,i}^{-1}$. Thus, the following equation holds:

\vspace{-15pt}

\begin{equation}
\begin{aligned}
 w_{k,k-1} \times \dots \times w_{2,1} &= w_{k-1,k}^{-1} \times \dots \times w_{1,2}^{-1}\\
                                       &= (w_{1,2} \times \dots \times w_{k-1,k})^{-1}\\
                                       &> 1.
\nonumber
\end{aligned}
\vspace{-15pt}
\end{equation}
\vspace{-5pt}
\end{proof}

It is almost impossible to have assets on different pools with exactly equal prices. Moreover, note that we are not considering transaction and swap fees. We make sure that the arbitrage revenue covers the costs in the validation step.



\noindent
\textbf{Transaction Ordering.} 
We introduce transaction ordering to enlarge the profitability of the existing arbitrage opportunities and even construct new arbitrage opportunities.
We optimize the trade prices of the assets (i.e., increase the weight of edge) involved in the cycle by transaction ordering for each cycle separately
. Given $path = [c_1\xrightarrow{w_{1,2}}c_2 \dots c_{k-1}\xrightarrow{w_{k-1,k}}{c_{k}}]$, we prioritises the execution of all transactions $T_{c_{i+1}\leftarrow c_{i}}$, where $i$ is an integer and $i \in [1, k-1]$, in mempool, which increase the price $w_{i,i+1}$ of swapping $c_{i}$ for $c_{i+1}$ when executing the $path$. As a result, we can end up with more $c_{1}$.




\vspace{1pt}
\noindent
\textbf{Trading Amount Optimization and Strategy Determination.}
For a given path, we need to determine the amount of trading assets for each action in the path. If the amount is inappropriate, the arbitrage profit extracted through that path will be reduced. Especially for arbitrage paths, we default that each action will take the output of the previous action as input. Therefore, we only need to determine the amount for the first trading asset. 
We focus on arbitraging on constant product automatic market makers. Consider a liquidity pool of assets $c_0$ and $c_1$ with fee $f$. The amount of assets $c_0$ and $c_1$ reserved in the liquidity pool are $Q_0$ and $Q_1$, respectively. When a liquidity taker swaps $\Delta_0$ amount of asset $c_0$ for $\Delta_1$ amount of asset $c_1$, the constant product AMM specifies $(Q_0 + (1-f)\Delta_0)(Q_1 - \Delta_1) = Q_0Q_1$.
Assume that we have found an arbitrage path $[c_1\xrightarrow{w_{1,2}}c_2 \dots c_{k-1}\xrightarrow{w_{k-1,k}}{c_{k}}]$. To find the optimal amount $\Delta_1$ of assets $c_1$ sold, we can construct a virtual liquidity pool $\mathrm{L}_{v}$ consisting of assets $c_1$ and $c_k$. By iterating Equation~\eqref{eqvirturelp}
, we can obtain the reserved amounts $Q_{v_0}$ and $Q_{v_0}'$ of assets $c_1$ and $c_k$ in $\mathrm{L}_{v}$~\cite{arbitrageanalysis2020}. 
\vspace{-3pt}
\begin{equation}
\label{eqvirturelp}
\begin{aligned}
& Q_{v_0} = \frac{Q_{v_0}Q_i'}{Q_i'+Q_{v_0}'(1-f)}, \\
& Q_{v_0}' = \frac{(1-f)Q_{v_0}'Q_{i+1}}{Q_i'+Q_{v_0}'(1-f)}.\\
\end{aligned}
\vspace{-3pt}
\end{equation}
As a rusult, the optimal amount $\Delta_1 = \frac{\sqrt{Q_{v_0}Q_{v_0}'(1-f)}-Q_{v_0}}{1-f}$.
For other types of AMMs (e.g., Bancor~\cite{hertzog2017bancor}), the optimal amount can be found by gradually increasing the amount of $c_1$ sold until the revenue no longer increases. 

We obtain the corresponding arbitrage strategy after performing transaction ordering and trading amount optimization for each cycle. 
We greedily select the most profitable arbitrage strategy from all possible arbitrage strategies for each round. 

\vspace{1pt}
\noindent
\textbf{Validation and State Updating.}
For the arbitrage strategy we find, we first execute it locally to confirm that the revenue is sufficient to cover the transaction and swap fees. 
If profitable, we include the transactions into the block in the order specified by the arbitrage strategy. 
Moreover, 
the price of assets changes dynamically with the trading volume in the DeFi AMM platform. 
Therefore, our algorithm needs to update the state $s$ and the graph $g$ after each execution of an arbitrage strategy to accommodate dynamic price changes.
 


\begin{algorithm} [ht]
\DontPrintSemicolon
\SetAlgoNoEnd 
\KwIn{$s_0 \leftarrow$ initial state, $target \leftarrow$ Minumum revenue target}
\KwOut{Arbitrage strategy sequence $\mathbf{strategy}$}
$s \leftarrow s_0$; $g \leftarrow buildGraph(\mathbf{V}, \mathbf{E}, s)$; $strategy \leftarrow None$\;  \label{alg1:line1}
\While{true}{
    $bestPath, bestAmount, bestT_{others} \leftarrow None$\;
    $bestRevenue \leftarrow 0$\;
    $\mathbf{path} \leftarrow getCycles(g)$\; \label{alg1:line2}
    \For{each $path \in \mathbf{path}$}{
        $T_{others}, s_{tmp} \leftarrow orderTxs(path, s)$\; \label{alg1:line3}
        $amount, revenue \leftarrow search(path, s_{tmp})$\; \label{alg1:line4}
        \If{$revenue > target$ and $revenue > bestRevenue$}{ \label{alg1:line5:start}
            $bestPath \leftarrow path$\;
            $bestAmount \leftarrow amount$\;
            $bestT_{others} \leftarrow T_{others}$\;
            $bestRevenue \leftarrow revenue$\; \label{alg1:line5:end}
        }
    }
    \eIf{$bestPath$ in not $None$}{ 
        $strategy \leftarrow (bestT_{others}, bestPath, bestAmount)$\;
        $\mathbf{strategy} \leftarrow \mathbf{strategy} \cup \{strategy\} $\;
        $s \leftarrow execute(strategy, s)$\; \label{alg1:line6:start}
        $g \leftarrow {buildGraph}(\mathbf{V}, \mathbf{E}, s)$\; \label{alg1:line6:end}
    }{
        break\;
    }
}
\Return $\mathbf{strategy}$\;
\;
\SetKwFunction{FuncA}{$buildGraph$}
\SetKwProg{Fn}{Function}{:}{end}
\Fn{\FuncA{$\mathbf{V}, \mathbf{E}, s$}}{
    Find the liquidity pool with the best price for each $e \in \mathbf{E}$\;
    Build the weighted directed graph $g$, where $w_{i,j} = \frac{Q_{c_j}}{Q_{c_i}}$\;
    \Return{$g$}\;
}
\SetKwFunction{FuncC}{$execute$}
\Fn{\FuncC{$strategy, s$}}{
    $s' \leftarrow$ State after the execution of $strategy$\;
    \Return{$s'$}\;
}
\SetKwFunction{FuncB}{$getCycles$}
\Fn{\FuncB{$g$}}{
    $\mathbf{cycles} \leftarrow DFS(g)$\;
    $\mathbf{path} \leftarrow None$\;
    \For{each cycle $\in \mathbf{cycles}$}{
        $path \leftarrow$ connect $\mathbb{T}$'s base cryptocurrency asset with $cycle$\;
        $\mathbf{path} \leftarrow \mathbf{path} \cup 
        \{ path \} $\;
    }
    \Return{$\mathbf{path}$}\;
}
\caption{Delayed Transaction Ordering Algorithm}
\label{al:DTOA}
\end{algorithm}

The DTOA process is shown in Algorithm~\ref{al:DTOA} and the following steps are repeated until no new arbitrage strategy is discovered: (i) build graph $g$ based on the system state $s$ (Line~\ref{alg1:line1}); (ii) detect all cycles (DFS) in the graph $g$ and construct paths $\mathbf{path}$ (Line~\ref{alg1:line2}); (iii) order transactions based on each path $path \in \mathbf{path}$ separately to optimize the trading price of the assets (Line~\ref{alg1:line3}); (iv) determine the optimal trading amounts for each path $path$ (Line~\ref{alg1:line4}); (v) determine the optimal path $path$ (Line~\ref{alg1:line5:start} to~\ref{alg1:line5:end}); (vi) execute the arbitrage strategy and update the state $s$ (Line~\ref{alg1:line6:start} to~\ref{alg1:line6:end}).
\vspace{-3pt}
\section{Fair and Automated Bribery}
\vspace{-3pt}
\label{Implement of Bribery Attack}
\label{sec:Implement of Bribery Attack}



In this section, we present our bribery smart contract that ensures trustless bribery fairness, and our bribery client that automates the bribery and arbitrage process.

\vspace{-6pt} 
\subsection{Bribery Smart Contract}
\vspace{-3pt}
\label{subsec:briberySC}

The bribery smart contract is designed to guarantee the fairness of bribery and enables the briber and bribee to complete the bribery process without mutual trust. When the bribee completes the appropriate action as required (e.g., voting the briber's block), the bribee is indeed paid through the smart contract. In turn, the smart contract verifies the evidence submitted by the bribee to ensure that the bribery fee is only paid to the bribee who takes the corresponding action.
To meet the above requirements, we design the bribery smart contract with the main functions: Attack Activation and Bribee Action Verification for bribers, and Bribee Withdrawal for bribees. 

\vspace{0pt} 
\noindent
\textbf{Attack Activation.} 
The bribery contract provides the bribers with the \texttt{activate()} function to initiate the bribery attack.
The \texttt{activate()} function includes setting the bribery state to $\mathbf{True}$, 
setting the attack time to slot $t$, determining the bribery target $\chi \in {0,1}$ based on the adaptive bribery strategy ($\chi =0$ represents bribing validators, $\chi = 1$ represents bribing proposers), and setting the bribery fee $\epsilon$ (the bribery fee must satisfy Equation~\eqref{bribery_fee1} when bribing validators, and must exceed the block reward $R_{P}$ when bribing proposers), transferring  bribery funds (which must exceed the total cost $\tau$) into the bribery contract and triggering the bribery event $BriberyEvent(t,\chi,\epsilon)$. Note that if the bribery contract has not yet been deployed on Ethereum, the adversary (bribery client) initiates a transaction, $T_{create}$, to deploy the bribery contract. 

\vspace{0pt} 
\noindent
\textbf{Bribee Action Verification.}
The bribery contract verify the action of bribees (bribed proposers and validators) to ensure that only those who perform the corresponding actions can receive the bribe.
For bribed proposers, the briber's block will only be valid and the bribery contract will be activated, allowing the bribed proposer to withdraw funds from the bribery contract, when the proposer performs the corresponding action (delaying block production). Therefore, the contract does not need to verify the action of the proposer.

However, validators may forge votes to steal bribery fee, so we need to verify whether validators vote to delay blocks.
Unfortunately, smart contracts cannot directly verify whether a given vote has been recorded on the chain in Ethereum 2.0. In this work, we accomplish the verification task with the help of the Chainlink Oracle~\cite{breidenbach2021chainlink}. The verification process is as follows: (i) The \texttt{acceptBribe()} function sends a request to the Oracle contract with the parameters (such as the hash of the vote and the block number containing the vote) provided by the bribee; (ii) The Oracle contract writes the request information to the Ethereum event log. (iii) The Chainlink node subscribing to the event gets the request information from the log. (iv) The Chainlink node uses the Beacon API \texttt{Beacon.get\_block\_attestations} to get the votes (attestations) in the corresponding block and verifies the existence of the bribed validator's vote. (v) The Chainlink node writes the verification result to the Oracle contract. (vi) The Oracle contract returns the result to the bribery contract.

\vspace{0pt} 
\noindent
\textbf{Bribee Withdrawal.}
The bribery contract provides the bribees with the \texttt{acceptBribe()} function to extract the bribery fee.
For bribed proposers, they only need to initiate a transaction in any slots after accepting the bribe to call the \texttt{acceptBribe()} function to withdraw the bribery fee.
For the bribed validators, they need to call the \texttt{acceptBribe()} function  after observing their vote being included in a block.
The \texttt{acceptBribe()} function requires the bribed validators to the correct validator address signature, the block number containing the validator's vote, and the hash of the vote. The bribery contract verifies the existence of the bribee's vote for block $n+1$ within the corresponding block by the Chainlink Oracle. 
Any incorrect values will result in a failed bribee withdrawal. If the verification passes, the bribery contract sends the bribery fee to the bribee.

\vspace{0pt} 
\noindent
\textbf{Bribery Cost Reduction.} 
 The extraction of bribery fee by the bribee incurs specific transaction fee, which the briber should subsidize. However, transaction fee on the Ethereum mainnet are expensive, and numerous validators require subsidies. 
 Therefore, we propose a method to reduce the cost of bribery: deploy the bribery contract on Layer 2, where the extraction of bribery fee can be performed with significantly lower transaction costs than on Layer 1. Moreover, the bribees are required to extract bribery fee daily or weekly, reducing the number of transactions requiring subsidies.
Furthermore, the bribery smart contract can also be deployed on zk Layer 2 such as Scroll~\cite{scroll} to protect the privacy of both the briber and bribees.

\vspace{-6pt} 
\subsection{Bribery Client}
\vspace{-3pt}
\label{subsec:client}
We now discuss the functions that the bribery client provide to automate the execution of bribery and arbitrage. Proposers and validators are required to install our bribery client to join the bribery process.


The bribers contain malicious proposers and validators.
Malicious proposers use the bribery client to statistic the proportion of bribable validators (Client Type Statistic), decide whether and how to delay block production (Attack Decision) and produce delayed arbitrage block (Delayed Block Production). Malicious validators use the bribery client to vote for the delayed block produced by the malicious proposer (Bribery Voting) and delay voting (Delayed Voting). The bribees contain bribed proposers and validators. The bribee uses the bribery client to decide whether to accept the bribe (Bribery Acceptance Decision). Bribed proposers will give up the block proposing opportunity and delay block production for 4 seconds (Delayed Block Production), while bribed validators will vote for the delayed block (Bribery Voting).

\vspace{0pt} 
\noindent
\textbf{Client Type Statistic.} 
The bribery client counts the proportion of validators in the network that have installed the bribery client for subsequent decision of bribery strategy.
The clients report their client type in their node identification. 
These reports can be obtained through node metadata, protocol handshake information, or other network communication methods.

\vspace{0pt} 
\noindent
\textbf{Attack Decision.} 
The bribery client assists the adversary decide whether and how to delay block production based on the proportion of bribable validators and revenue of producing block normally. 
When the validator controlled by the adversary $\mathbb{T}$ is selected to be the proposer for slot $t$ (which is determined at the beginning of each epoch), the bribery client first tries to produce block normally and discovers arbitrage opportunities using DTOA algorithm. 
Suppose the proportion of rational validators who have installed the bribery client exceeds the minimum required proportion (as defined in Equation~\eqref{bribery_num1}) and the revenue from normal block production can cover the cost of bribing validators. In that case, the bribery client bribes the validators of the next slot.
Suppose there are not enough validators with bribery client, but the proposer of the next slot has installed the bribery client and the revenue from normal block production can cover the cost of bribing proposers. In that case, the bribery client bribes the proposer of the next slot. The client activates the bribery contract by calling the \texttt{activate()} function to initiate a bribery attack.
If none of the above conditions are met, the bribery client produces blocks normally.


\vspace{3pt}
\noindent
\textbf{Delayed Block Production.} 
The bribery client delays producing the block based on the result of the DTOA algorithm.
Considering the running time required by the DTOA algorithm, the malicious proposer $\mathrm{P}_{t}$ is required to run the DTOA algorithm at the 8th second of slot $t$ to detect arbitrage opportunities and generate the corresponding arbitrage strategy sequence. The malicious proposer $\mathrm{P}_{t}$ delays the production block $n$ to the 0th second of slot $t+1$ and includes transactions in the order of transactions determined by the DTOA algorithm. If the block still has extra space, $\mathrm{P}_{t}$ also includes other transactions with high gas prices into block $n$. Moreover, bribery client $\mathrm{P}_{t}$ forges the timestamp of block $n$ to be the timestamp during the slot $t$ and publishes it at the 0th second of slot $t$. 
As a result of the bribery attack, the validators of slot $t+1$ attest block $n$. 

We choose to start running the algorithm at the 8th second of slot $t$ mainly to balance information integrity and performance requirements.
On the one hand, rational validators of slot $t$ will vote and broadcast their vote at the 4th second of slot $t$. At the 8th second of slot $t$, proposer $\mathrm{P}_{t}$ can receive most of the validators' votes~\cite{gossipsub2020}. This means that proposer $\mathrm{P}_{t}$'s view of the chain is closer to the truth. 
On the other hand, compared to producing blocks normally, an additional 8 seconds of transactions 
are accumulated in the mempool. 
These newly accumulated transactions will be used to enhance the arbitrage profits.
If the waiting time is too short, proposer $\mathrm{P}_{t}$ cannot collect sufficient transactions and may produce an invalid block with a wrong view due to incomplete votes.
If $\mathrm{P}_{t}$ waits longer, while more transactions can be accumulated, the algorithm may not be able to complete before slot $t+1$, thus affecting block production.




\vspace{0pt} 
\noindent
\textbf{Bribery Acceptance Decision.} 
Rational validators and proposers with bribery clients can listen to bribery events and decide whether to accept the bribe. When they discover the bribery event message from the briber, they check if the bribery fee satisfies the minimal required fee, i.e., Equation~\eqref{bribery_fee1} for validators or greater than $R_P$ for proposers. If it does, validators accept the bribe and vote for the briber's block $n$, proposers delay block production for 4 seconds. Otherwise, validators follow the fork choice rule to vote for the block $n+1$,  proposers produce the block normally. 

\vspace{0pt} 
\noindent
\textbf{Bribery Voting.} 
The bribed validators and malicious validators with bribery client can ignore the fork choice rule to vote for the briber's block. 

\vspace{0pt} 
\noindent
\textbf{Delayed Voting.} 
To collect enough votes, malicious validators of slot $t$ can delay the voting process to vote for block $n$ once they receive the block $n$ produced by $\mathrm{P}_{t}$. 

\vspace{-5pt}
\section{Implementation and Evaluation}
\vspace{-1pt}

\subsection{Implementation and Experimental Setup}
\vspace{-3pt}
We implement a prototype of our bribery client (Section~\ref{subsec:client} for detailed designs) based on \textit{prysm}~\cite{prysm}, a Golang implementation of the Ethereum 2.0 client. 
We develop a bribery smart contract (see Section~\ref{subsec:briberySC} for detailed designs) based on solidity 0.8.0 version. 
To complete the verification function, we register the corresponding Job for Chainlink node, which is written in python. 
We build a local Ethereum 2.0 network with 64 validators on Amazon t3a.2xlarge instances (with 8 AMD EPYC vCPU, 32GB RAM). 
This network also contains an \textit{Optimism}~\cite{optimism} Layer 2 network, on which we deploy our bribery smart contract to reduce the bribery cost.


By default, we select 26 assets and 100 actions from Uniswap and SushiSwap (see Table~\ref{tab:liquidity pools}), which involves 50 liquidity pools, based on our heuristic liquidity pool selection. 
Each asset is available for trading on Uniswap and SushiSwap. 
Both adversaries $\mathbb{T}$ and $\mathbb{B}$ respectively control 20\% of validators if not specified. 
The adversary $\mathbb{T}$ runs BriDe Arbitrager to boost its profits. 
The adversary $\mathbb{T}$ performs a bribery attack based on the adaptive bribery strategy in Section~\ref{section: optimal Bribe} and pays 1 Gwei bribery fee (which can be any value greater than 0) to each bribed validators. 

To accurately evaluate the profit of BriDe Arbitrager, we calculate the total bribery fee based on the total validators' number in the Ethereum of 600,000~\cite{beaconcha}.
We validate the BriDe Arbitrager on Ethereum data from block 12,000,000 to 12,100,000. We query the state of the 50 liquidity pools at each block through the subgraph service~\cite{unisubgrapg2021, sushisubgrapg2023}.

\begin{table}[tbp]
  \centering
  \caption{Liquidity pools on Uniswap and SushiSwap selected by the heuristic liquidity pool selection strategy}
  \vspace{-3pt}
  \begin{tabular}{lll}
    \toprule
    Liquidity Pool  &  Liquidity Pool  &  Liquidity Pool \\
    \midrule
    ETH $\leftrightarrows$ SFI & ETH $\leftrightarrows$  GRT & ETH $\leftrightarrows$  SNK\\
    ETH $\leftrightarrows$  SUSHI & ETH $\leftrightarrows$ FXS & ETH $\leftrightarrows$  OCEANK\\
    ETH $\leftrightarrows$  DAI & ETH $\leftrightarrows$  RFOX & ETH $\leftrightarrows$  PNK\\
    ETH $\leftrightarrows$ UNI & ETH $\leftrightarrows$  COMP & ETH $\leftrightarrows$  INV\\
    ETH $\leftrightarrows$  AMPL & ETH $\leftrightarrows$ LINK & ETH $\leftrightarrows$  MKR\\
    ETH $\leftrightarrows$  USDT & ETH $\leftrightarrows$  WBTC & ETH $\leftrightarrows$  DPI\\
    ETH $\leftrightarrows$ ALCX & ETH $\leftrightarrows$  AAVE & ETH $\leftrightarrows$  REVV \\
    ETH $\leftrightarrows$  LDO & ETH $\leftrightarrows$  YFI\\
    ETH $\leftrightarrows$  CRV & ETH $\leftrightarrows$  ANT \\
    \bottomrule
  \end{tabular}
  \label{tab:liquidity pools}
\end{table}
\vspace{-3pt}

\vspace{-7pt}
\subsection{Profits of BriDe Arbitrager}
\vspace{-3pt}
\label{subsec:Profits of Bribery-MEV Extractor}
We compare the cumulative profits (i.e., revenue minus cost) of BriDe Arbitrager run by $\mathbb{T}$ with the following four baseline arbitrage detection algorithms 
: (i) Negative cycle detection~\cite{zhou2021just} (NCD); (ii) Delayed transaction ordering based on NCD (DTO-NCD); (iii) Arbitrage detection with DFS~\cite{dfsArbitrage} (DFS); (iv) Delayed transaction ordering based on optimal DFS (DTO-DFS). 
For all the arbitrage opportunities found by the above algorithms, we use the trading amount optimization mentioned in Section~\ref{Delayed Arbitrage Arbitrage Details} to determine the optimal trading amount. 

\vspace{1pt} 
\noindent
\textbf{NCD.} 
This represents the previous arbitrage detection: a negative cycle detection algorithm~\cite{zhou2021just} using Bellman-Ford-Moore
algorithm to find negative cycles. 
Whenever it finds a negative cycle, it extracts the profit.

\vspace{1pt} 
\noindent
\textbf{DTO-NCD.}
This requires adversaries to delay block production via bribery attacks to accumulate more transactions (order flow). 
Then, for all negative cycles found by NCD, 
we apply transaction ordering (ordering more transactions in the extended period) to increase the trading prices of the assets, i.e., the weight of the edge, involved in the negative cycles. 
We greedily select the negative cycle with the largest revenue based on the optimized prices. 

\vspace{1pt} 
\noindent
\textbf{DFS.} 
This is the arbitrage detection algorithm using DFS. In this algorithm, 
we use DFS to search for all possible cycles and then greedily choose the cycle with the largest revenue. 

\vspace{1pt} 
\noindent
\textbf{DTO-DFS.} 
This algorithm is similar to our BriDe Arbitrager.
However, in DTO-DFS, we first use DFS to find the most profitable cycle and then order the transactions on the most profitable cycle.
In contrast, in BriDe Arbitrager, we first order the transactions on all cycles we find and then select the most profitable cycle.

\begin{figure}[tp]
\centerline{\includegraphics[width=0.35\textwidth]{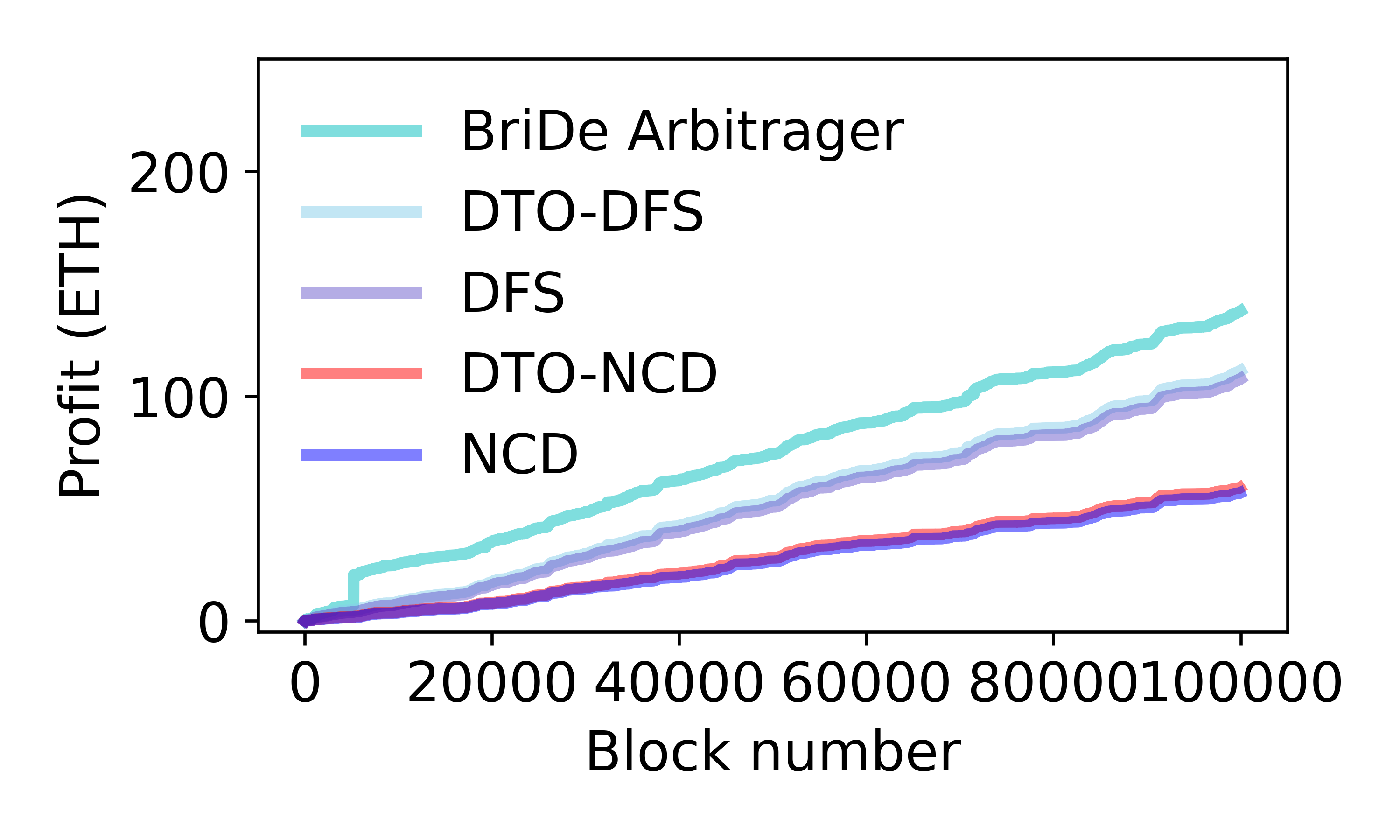}}
\vspace{-13pt}
\caption{Profits of different arbitrage algorithms.}
\label{result:profits camparision}
\vspace{-14pt}
\end{figure}

The results in Figure~\ref{result:profits camparision} show that BriDe Arbitrager has greater profitability than all other arbitrage detection algorithms, with a daily profit of 8.66\,ETH (16,442.23\,USD). Compared to the NCD and DFS, BriDe Arbitrager can boost profits by at least more than 30\,ETH. 
The most significant profitability improvement of the BriDe Arbitrager comes from its ability to build new arbitrage opportunities by increasing the price of specific assets through DTOA, where an 8 second delay allows BriDe Arbitrager to accumulate more transactions.
In other words, BriDe Arbitrager can identify arbitrage opportunities that may arise from transactions of other users that have not been executed.
While algorithms (i) and (iii) attempt to discover arbitrage opportunities that exist in the current state, algorithms (ii) and (iv) attempt to amplify existing arbitrage opportunities by taking advantage of transactions from other users that remain unexecuted. 
For instance, by prioritising executing these transactions that swap SFI for ETH on Uniswap or swap ETH for DAI on SushiSwap, and then executing the arbitrage transactions $[ETH\xrightarrow{Uniswap}SFI\xrightarrow{SushiSwap}ETH]$ and $[ETH\xrightarrow{Uniswap}DAI\xrightarrow{SushiSwap}ETH]$ at block 12,005,217, the BriDe Arbitrager ends up with a profit of 13.39\,ETH (25,432.97\,USD). 
However, there is no arbitrage opportunity for the above paths at the beginning of block 12,005,217. 
Therefore, other arbitrage detection algorithms cannot extract the above profit.

\begin{figure}[tp]
  \centering
  \begin{subfigure}[t]{0.235\textwidth}
    \centering
    \includegraphics[width=\textwidth]{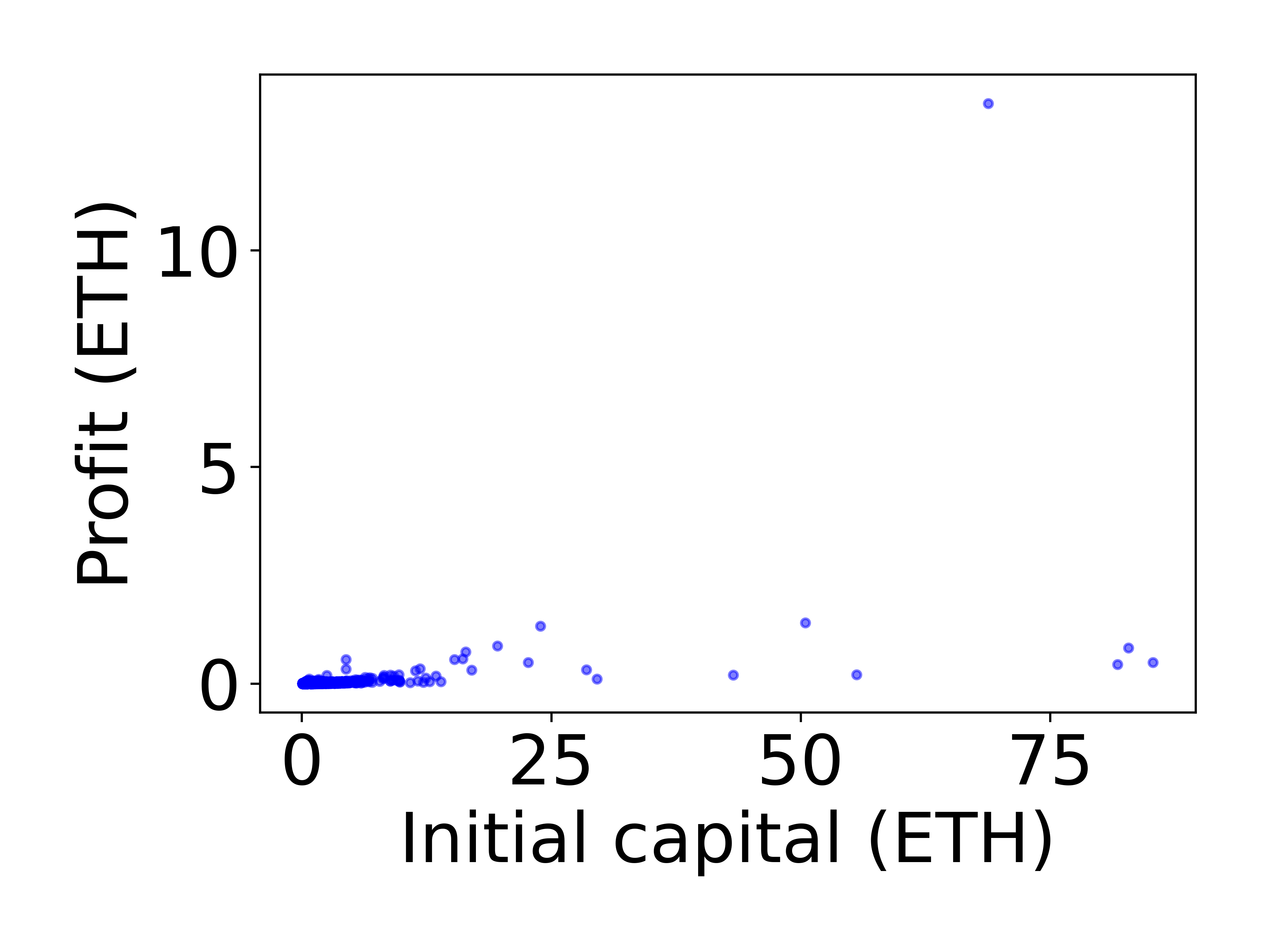}
    \captionsetup{justification=centering}
     \vspace{-21pt}
    \caption{BriDe Arbitrager w/o \\ flash loans.}
    \label{fig:subfig1}
  \end{subfigure}
  \begin{subfigure}[t]{0.235\textwidth}
    \centering
    \includegraphics[width=\textwidth]{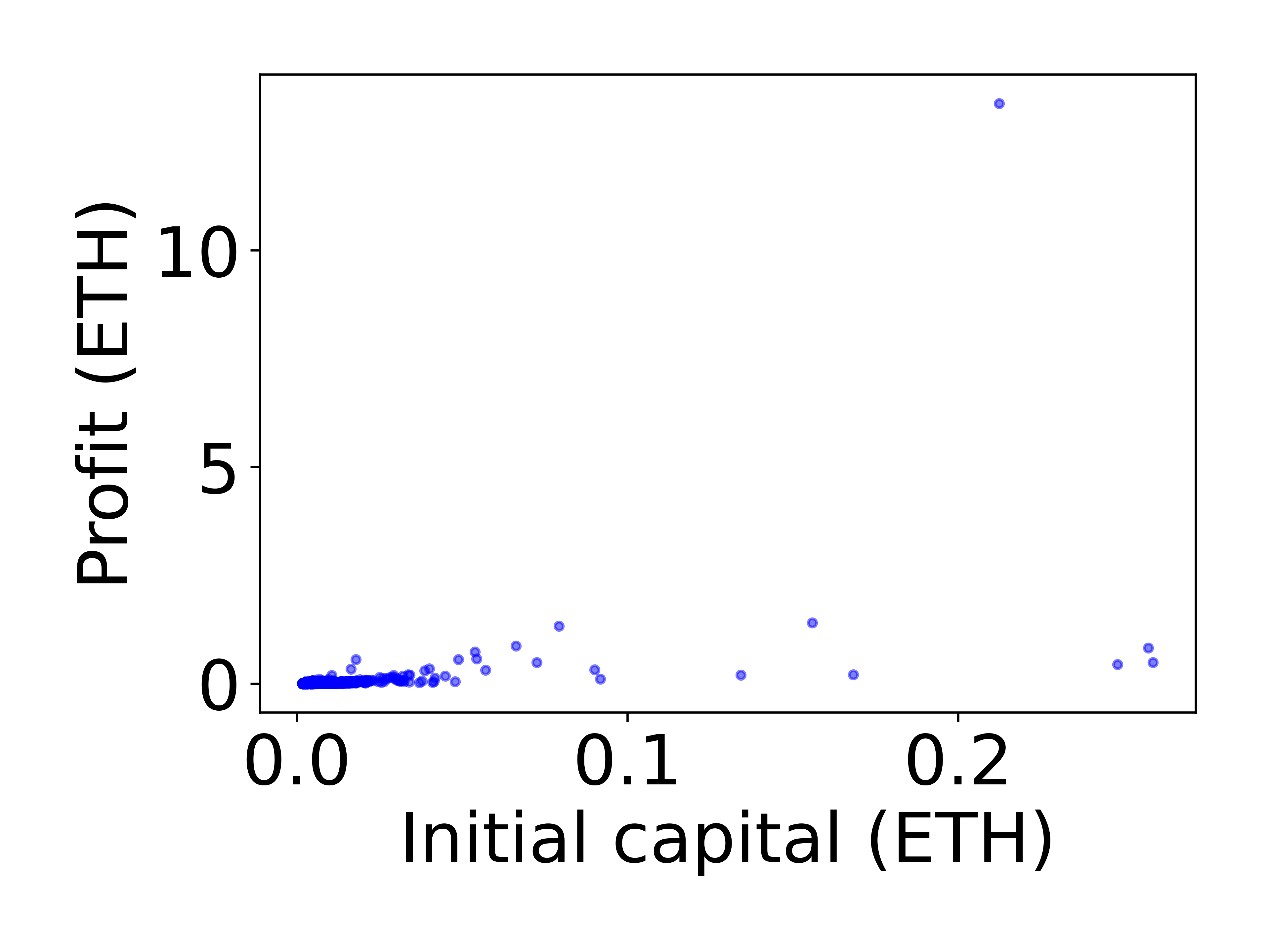}
    \captionsetup{justification=centering}
    \vspace{-21pt}
    \caption{BriDe Arbitrager with \\ flash loans.}
    \label{fig:subfig2}
  \end{subfigure}
    \captionsetup{justification=centering}
    \vspace{-5pt}
  \caption{Profit as a function of the initial capital \\  with and without flash loans.}
  \label{result:capital with and without flash loans}
  \vspace{-12pt}
\end{figure}

We represent the profit of each arbitrage strategy sequence in a block found by BriDe Arbitrager against the initial capital required by that arbitrage strategy sequence in Figure~\ref{result:capital with and without flash loans}. 
Only 9 out of 12,230 arbitrage strategy sequences require an initial capital of more than 25\,ETH. 
Most of the arbitrage strategy sequences have profits below 2\,ETH. 
If flash loans are used, the initial capital required by the arbitrage strategy sequences will be reduced to less than 0.3\,ETH.

We visualize the profit distribution of BriDe Arbitrager in Figure~\ref{result:revenue distribution}. BriDe Arbitrager found 12,230 arbitrage strategy sequences that yielded profits of 137.83\,ETH (261,794.30\,USD). The most profitable arbitrage strategy sequence consists of 4 arbitrage strategies that yielded a profit of 13.39\,ETH (25,432.97\,USD). In general, the profit increases as more arbitrage strategies are included in the arbitrage strategy sequence.

\vspace{-5pt}
\subsection{Execution Time of DTOA}
\vspace{-3pt}

\begin{figure}[tp]
  \centering
  \begin{minipage}[t]{0.235\textwidth}
    \centering
    \includegraphics[width=\textwidth]{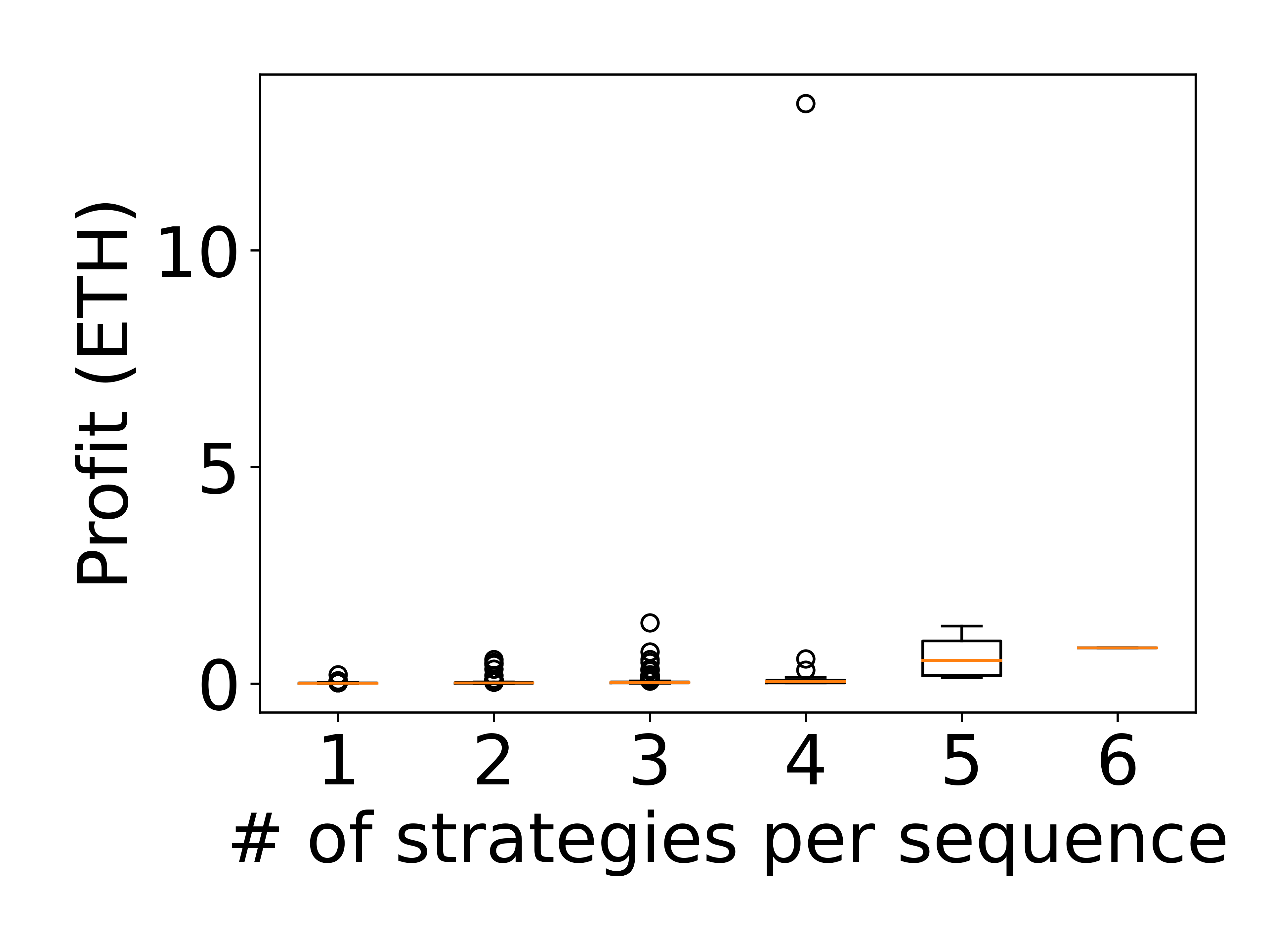}
    \captionsetup{justification=centering}
     \vspace{-15pt}
    \caption{Profit distribution.}
    \label{result:revenue distribution}
  \end{minipage}
  \hfill
  \begin{minipage}[t]{0.235\textwidth}
    \centering
    \captionsetup{justification=centering}
    \includegraphics[width=\textwidth]{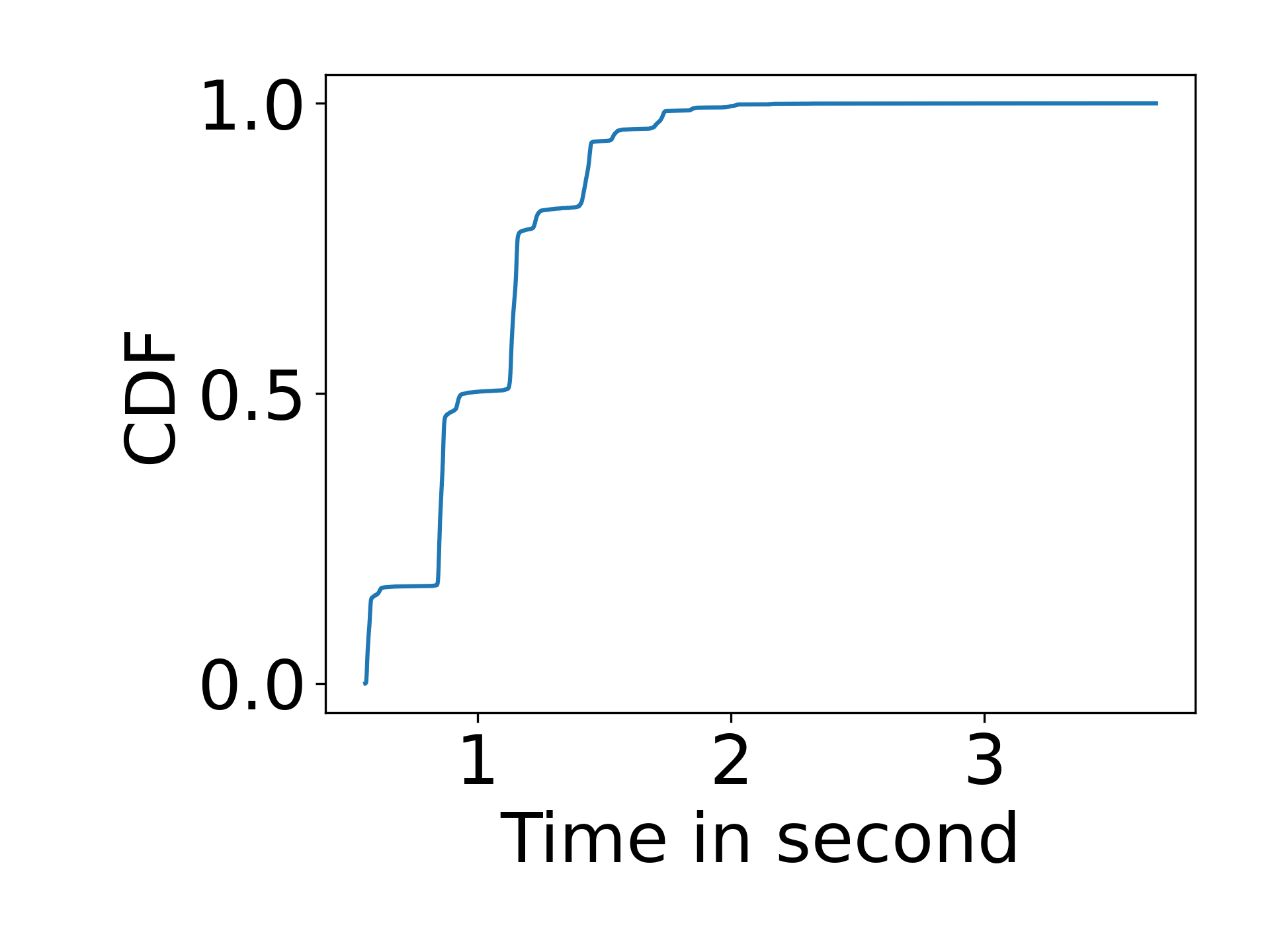}
     \vspace{-15pt}
    \caption{Time distribution.}
  \label{result:time distribution}
  \end{minipage}
\vspace{-10pt}
\end{figure}

We illustrate the execution time distribution of DTOA in Figure~\ref{result:time distribution}. For each new slot, the average execution time of DTOA is 1.03 seconds. Recalling the BriDe Arbitrager process, 
the DTOA algorithm is required to obtain the arbitrage strategy sequence in 4 seconds. The result shows that all arbitrage strategy sequences are found within the time limit of 4 seconds.

\begin{figure}[tp]
  \centering
  \begin{minipage}[t]{0.235\textwidth}
    \centering
    \includegraphics[width=\textwidth]{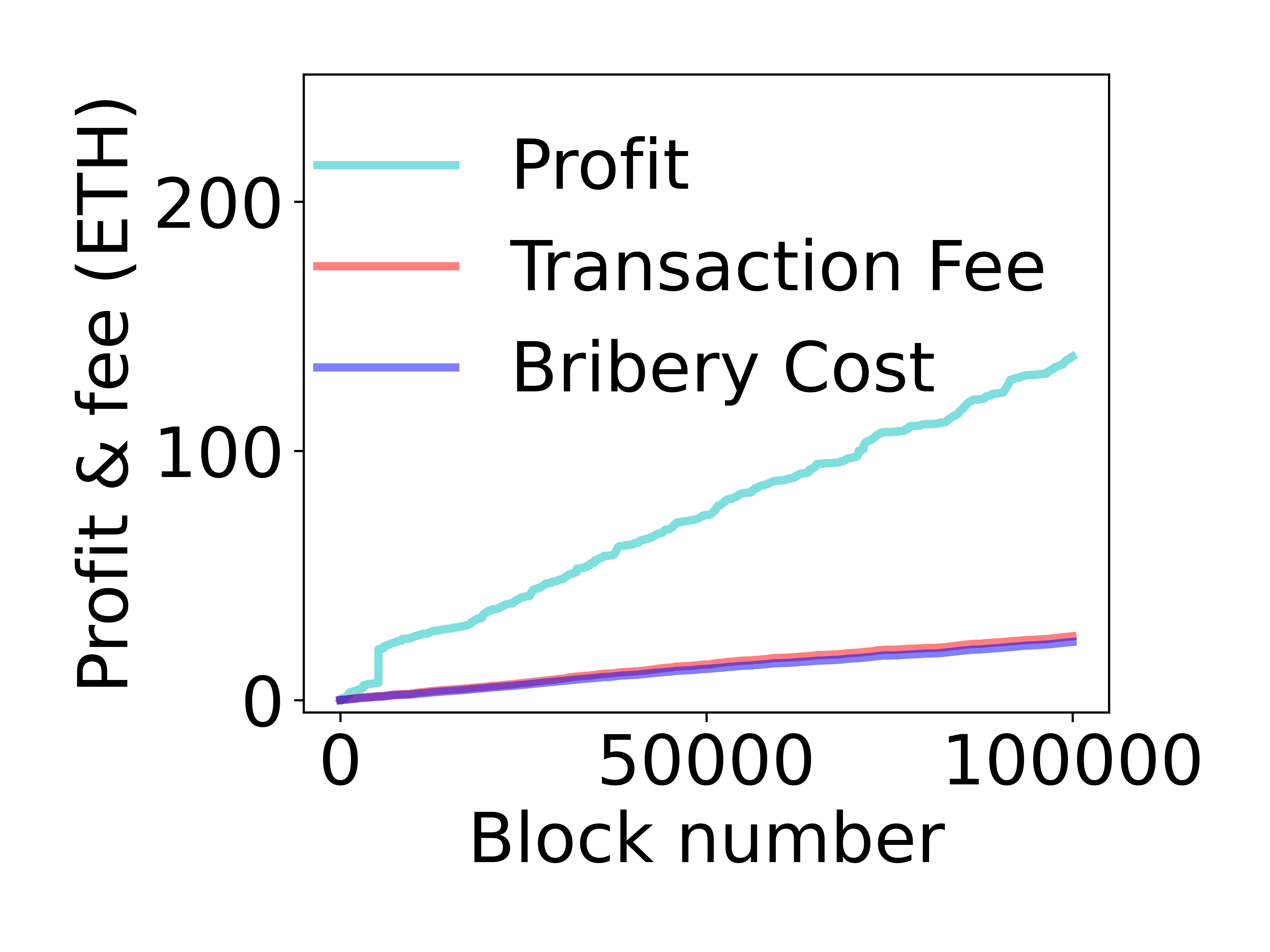}
    \captionsetup{justification=centering}
    \vspace{-20pt}
    \caption{Cumulative profits and cost.}
    \vspace{-10pt}
    \label{result:cumulative cost}
  \end{minipage}
  \hfill
  \begin{minipage}[t]{0.235\textwidth}
    \centering
    \includegraphics[width=\textwidth]{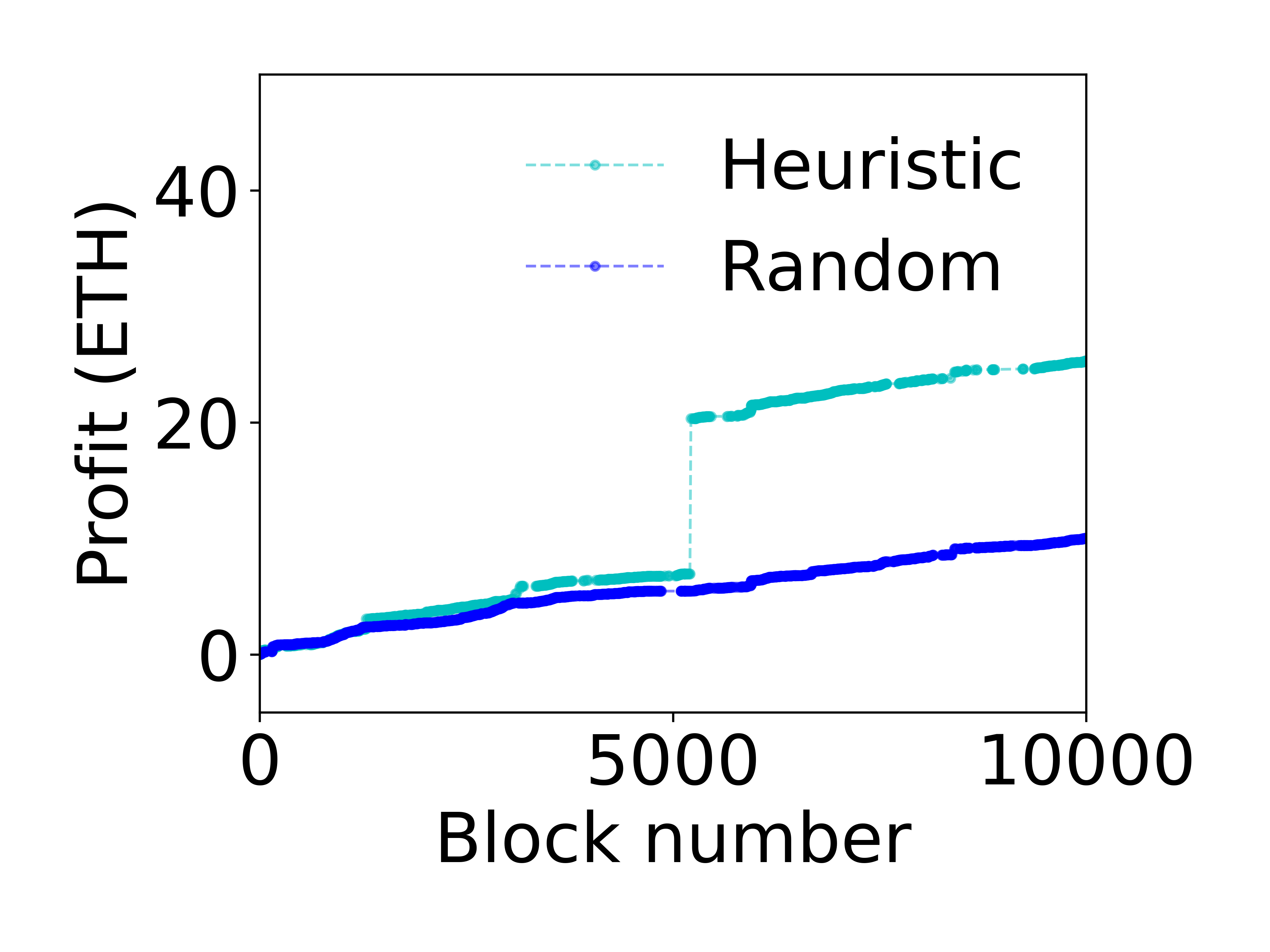}
    \vspace{-20pt}
    \caption{Profits of different liquidity pool selection strategies.}
    \label{result:comparison with random}
  \end{minipage}
  \vspace{-10pt}
\end{figure}



\vspace{-7pt}
\subsection{Cost of BriDe Arbitrager}
\vspace{-3pt}
The cost of BriDe Arbitrager consists of transaction fee and bribery cost. Figure~\ref{result:cumulative cost} shows the cumulative transaction fee, bribery cost and profit. Compared to profit, the cost of BriDe Arbitrager is small, accounting for 35.43\,\% of the profit.
Through the validation phase, we guarantee in the DTOA algorithm that the revenue of the discovered arbitrage strategy is sufficient to cover the transaction fees needed to execute it, which makes the profit positive.

\vspace{-7pt}
\subsection{Comparison with Random Liquidity Pool Selection}
\vspace{-3pt}
We compare the cumulative profits of our liquidity pool selection with that of randomly selected liquidity pools in Figure~\ref{result:comparison with random}. We select 20 liquidity pools based on the liquidity pool selection strategy and 20 liquidity pools based on the random strategy, respectively, from Table~\ref{tab:liquidity pools}.
To enable the cycle, we guarantee that the assets involved in the randomly selected liquidity pools can be traded on both Uniswap and SushiSwap. 
The result shows that applying BriDe Arbitrager in the liquidity pools selected according to our heuristic liquidity pool selection strategy can be more profitable for the same search size. 

\begin{figure}[tp]
  \centering
  \begin{minipage}[t]{0.235\textwidth}
    \centering
    \captionsetup{justification=centering}
    \includegraphics[width=\textwidth]{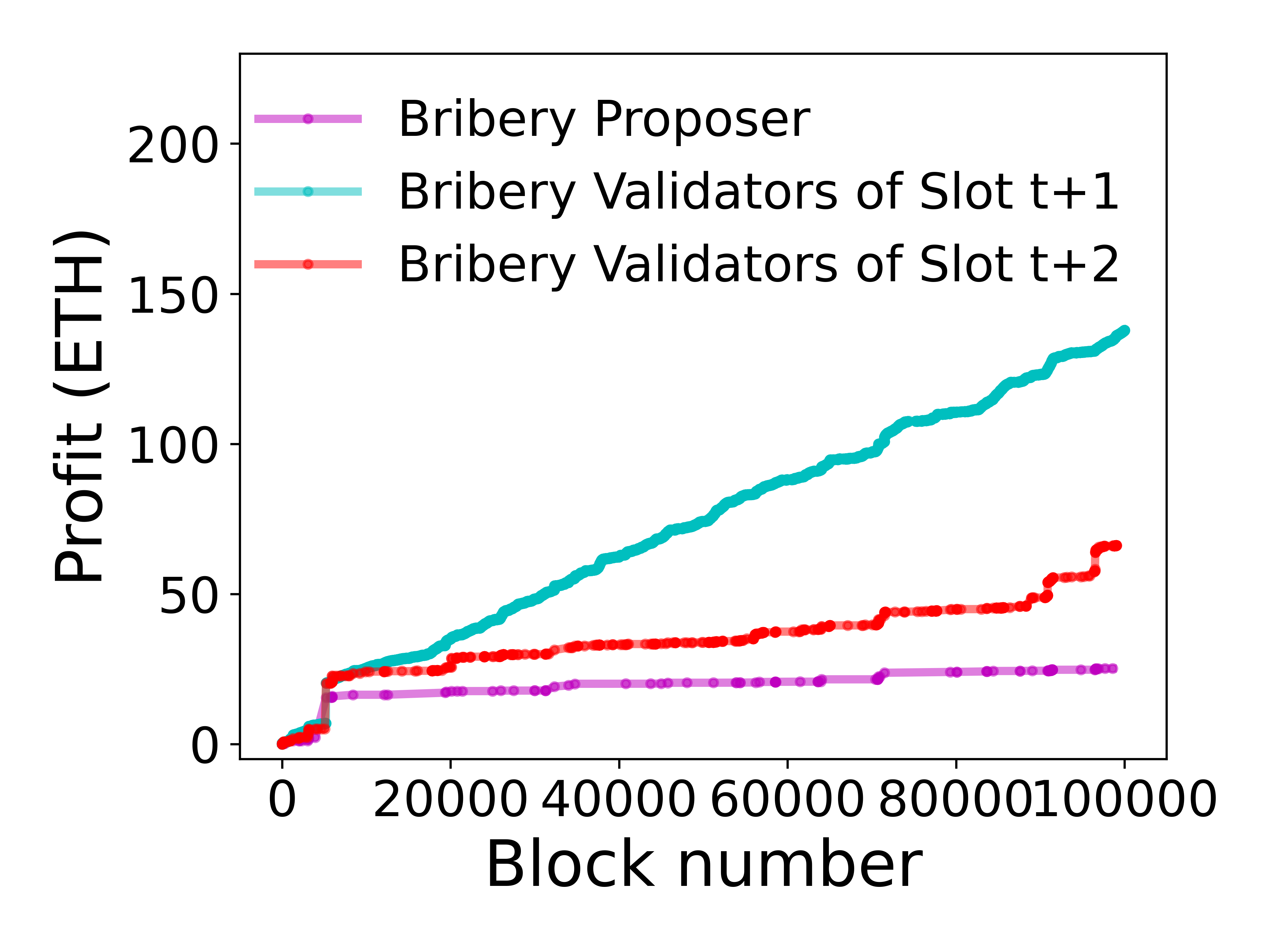}
    \captionsetup{justification=centering}
    \vspace{-15pt}
    \caption{Profits of different bribery strategies.}
    \vspace{-10pt}
    \label{result:comparison with other bribery}    
  \end{minipage}
  \hfill
  \begin{minipage}[t]{0.235\textwidth}
    \centering
    \captionsetup{justification=centering}
    \includegraphics[width=\textwidth]{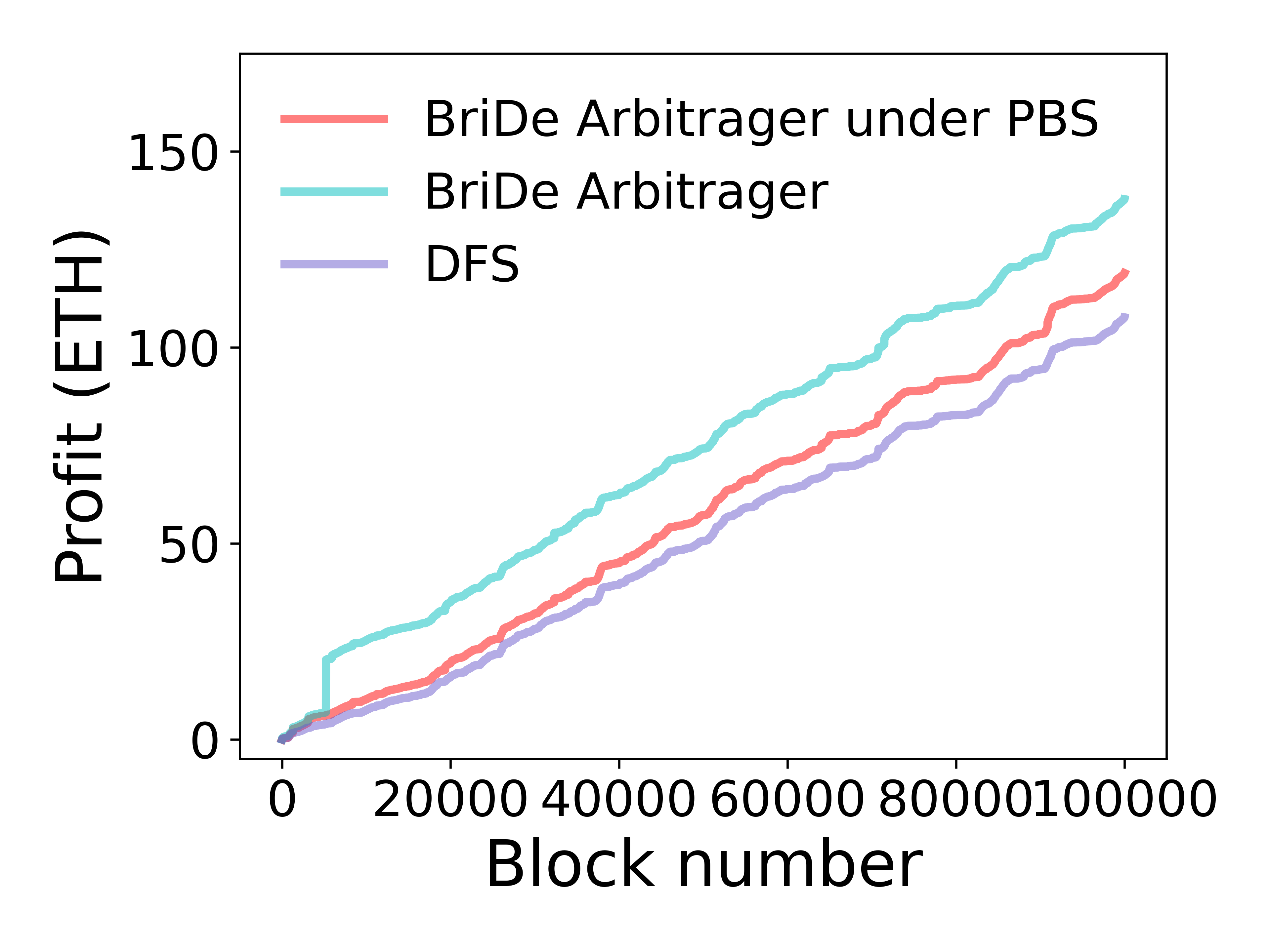}
    \vspace{-15pt}
    \caption{Profits under PBS.}
    \label{fig:pbs}
  \end{minipage}
\vspace{-10pt}
\end{figure}

\vspace{-7pt}
\subsection{Comparison with Other Bribery Strategies}
\vspace{-3pt}
Figure~\ref{result:comparison with other bribery} demonstrates the cumulative profits of three bribery strategies: (1) bribe proposer of slot $t+1$; (2) bribe validators of slot $t+1$; (3) bribe validators of slot $t+2$. The results show that the proposer $\mathrm{P_{t}}$ delays block production and bribes the validators of slot $t+1$ is the bribery strategy with maximum cumulative profit. 
As discussed in Section~\ref{section: optimal Bribe}, for the proposer $\mathrm{P_{t}}$, bribing the validator of slot $t+1$ has the same return as bribing the proposer of slot $t+1$, but the latter requires a higher cost and thus yields a lower profit. Although bribing validators of slot $t+2$ has a higher chance of discovering more high-quality arbitrage strategies for proposer $\mathrm{P_{t}}$, it needs to take a greater risk (i.e., pay more bribery fees).

\vspace{-7pt}
\subsection{Feasibility under Existing Ethereum}
\vspace{-3pt}
\label{subsec:feasibility}

Our bribery strategy is \emph{applicable for different proportions of malicious validators and rational validators.}
As derived and experimented by us, 
a malicious proposer who controls 20\% of validators needs to bribe an additional 20\% of validators of the next slot (attacked slot).
Correspondingly, a malicious proposer with less voting power needs to bribe more validators, and vice versa.
Under our designs, this is practical.

First, it is possible for an adversary to hold 20\% of the validators in the network. 
For example, a large entity Lido owns more than 20\% of the voting power~\cite{wahrstatter2023time}. 
Additionally, recall that validators join Ethereum to earn rewards. In the absence of bribery, validators adhere to the Ethereum protocol by honestly voting and producing blocks. When bribery exists, and the bribery fee satisfies Equation~\eqref{bribery_fee1}, rational validators with bribery clients can accept bribes to obtain the bribery fee. The expected profits of rational validators with bribery clients are greater than those without them. It makes sense that 20\% of validators can equip a bribery client to accept bribes since blockchain nodes are driven by profit. Under our adaptive bribery strategy, bribed rational validators earn an average of 225 Gwei/day more than honest validators.
Furthermore, our fully automated bribery client further promotes validators to accept bribes (Section \ref{subsec:client}).
Statistically, 37\% of existing validators use the go version of the Ethereum client 
~\cite{clidiversity}.
Therefore, it is also practical for them to simply migrate to our Golang version of bribery client.
Although we assume that all validators are rational, a malicious proposer can launch a bribery attack even in the existence of altruistic validators who always follow the blockchain protocol, as long as the proportion of bribable rational validators satisfies Equation~\eqref{bribery_num1}. 

More importantly, in the worst case, a malicious proposer can successfully delay block production, even if the voting power controlled by it and rational validators cannot meet the Equation~\eqref{bribery_num1}.
In this case, the malicious proposer needs to bribe the next slot's proposer instead of validators, which increases the bribery cost.

\begin{figure}[tp]
\centerline{\includegraphics[width=0.3\textwidth]{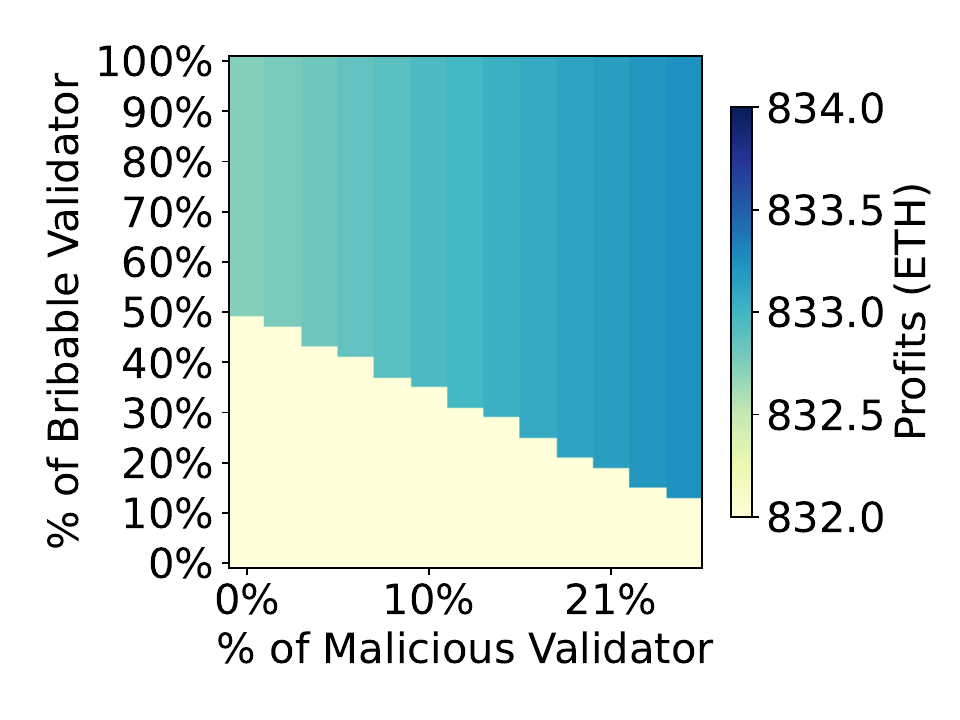}}
\vspace{-9pt}
\caption{Profits at different \% of malicious validators and bribable validators.}
\label{fig:profits at different validators}
\vspace{-20pt}
\end{figure}

We show the profit of BriDe Arbitrager at different proportions of malicious validators and bribable validators in Figure~\ref{fig:profits at different validators}. 
If the proportions of malicious validators and bribable rational validators satisfy Equation~\eqref{bribery_num1}, the malicious proposer ($\mathbb{T}$) bribes the next slot's validators, in which case $\mathbb{T}$ makes an average profit of at least 0.08\,ETH each time (the upper part of Figure ~\ref{fig:profits at different validators}). Otherwise, the malicious proposer bribes the proposer of the next slot (the lower part of Figure~\ref{fig:profits at different validators}). Since bribery costs of bribing the proposer of next slot three times as much as bribing the validators of next slot, $\mathbb{T}$ only profits an average of 0.012\,ETH each time.

\vspace{-3pt}
\vspace{-5pt}
\subsection{Feasibility under PBS}
\vspace{-3pt}
\label{Feasibility under PBS}
We explain the feasibility of BriDe Arbitrager under Proposer Builder Separation~\cite{pbs2021} in this subsection.

\vspace{3pt}
\noindent
\textbf{Extended Threat Model. }
We extend our threat model of Section~\ref{threat model} to consider the PBS mechanism. 
Under the PBS mechanism, Ethereum 2.0 splits block production into block building and proposing, with a proposer responsible for proposing blocks and a builder responsible for constructing blocks, i.e., determining the order of transactions within blocks.
Malicious proposer $\mathrm{P}_{t}$ utilizes the DTOA algorithm to identify an arbitrage strategy sequence and includes the corresponding transactions into the block in order. If there is still available space left in the block, the proposer selects as many transactions with high gas prices as possible from the mempool. These transactions are required not to impact the prices of assets in the liquidity pools associated with the arbitrage strategy sequence.
 The proposer then broadcasts the transactions involved in the arbitrage strategy sequence and the selected transactions from the mempool to the builders. Builders try to order the transactions selected by the proposer to maximize their profit. 
 Builders bid against each other to compete for the right to determine the transaction order for this slot and obtain corresponding profits.
 Builders send the sorted block and the corresponding bids back to the proposer. The proposer selects the block with the highest bid to obtain the corresponding bid.


\vspace{3pt}
\noindent
\textbf{Feasibility under PBS. }
The transactions that the proposer broadcasts to the builder are divided into three categories: the arbitrage transactions that we use to earn a profit (denoted $T_{arbitrage}$), the transactions selected by the DTOA algorithm to enhance arbitrage profits (denoted $T_{others}$), and unrelated transactions (denoted $T_{unrelated}$). The builder with the highest bid only can profit from transactions in $T_{others}$ and $T_{unrelated}$. The order of the transactions in $T_{unrelated}$ will be irrelevant to the proposer's revenue because it does not affect the liquidity pools involved in our arbitrage transactions. Therefore, we are only concerned with the order of transactions $T_{arbitrage}$ and $T_{others}$.

If the builder chooses to execute the transaction in $T_{others}$ before $T_{arbitrage}$, which is what we expect, we can get the expected revenue. However, if transactions in $T_{others}$ are executed after $T_{arbitrage}$, we may lose transaction fees because we are trading on a path without arbitrage opportunities. 
Since we cannot predict the actual behaviour of the builder, we simulate the sorting of builders by randomly determining the order of $T_{arbitrage}$ and $T_{others}$.
We present our experimental results in Figure~\ref{fig:pbs}, which indicate that although the profit of BriDe Arbitrager decreases under PBS, it is still more profitable than honestly producing blocks.
\vspace{-5pt}
\subsection{Feasibility under SSLE}
\label{Feasibility under SSLE}
\vspace{-3pt}
We explain the feasibility of BriDe Arbitrager under Secret non-Single Leader Election~\cite{ssle2022} (SSLE) in this section.

In the SSLE mechanism, the proposer for each slot is secretly elected. Their identity is not known to the entire network except for the proposers themselves.
The SSLE mechanism makes it impossible for us to determine whether the proposer of the next slot has installed the bribery client, thus we cannot judge whether delaying block production by bribing the next slot's proposer is feasible.
However, we can still obtain the client type of validators when communicating with them (SSLE does not hide validator identities), thus confirming the proportion of validators in the network who have installed bribery client. As long as the proportion of rational validators with bribery client installed meets the requirement, as in Equation~\eqref{bribery_num1}, malicious proposers can delay block production by bribing validators in the next slot. As discussed in Section~\ref{subsec:feasibility}, rational validators are profit-driven, so they all have an incentive to install the bribery client, meaning we will have a sufficient number of bribable validators. Therefore, the BriDe Arbitrager remains feasible under the SSLE mechanism.


\vspace{-6pt}
\section{Discussions}
\label{sec:limitation}



\noindent
\textbf{Long-term Bribery.}
In this work, we consider single-slot bribery. 
The malicious bribery proposer and the rational validators make decisions based on the current state, and the bribery attack lasts only a few time slots (i.e., one slot according to the analysis in Section~\ref{section: optimal Bribe}). However, a malicious bribing proposer can initiate long-term bribery attacks by continuously bribing validators over multiple slots to 
delay the production of malicious blocks more slots, and ensure the success of the bribery attack. 
Intuitively, long-term bribery has higher costs because we need to bribe more validators. However, the new arbitrage opportunities resulting from long-term bribery are insufficient to cover its increased cost. The trade-off between bribe timing and arbitrage presents an exciting direction for future research.

\vspace{3pt} 
\noindent
\textbf{Other MEV Detection Methods.}
The algorithm proposed in Section~\ref{sec:Delayed Arbitrage} focuses on identifying potential arbitrage opportunities and amplifying the profits from such arbitrage. However, it is essential to note that there are various forms of MEV opportunities in DeFi, such as liquidations, sandwich attacks, and more, all of which can yield significant profits for adversaries. In future work, we aim to design more comprehensive algorithms that can effectively leverage the benefits of bribery attacks to discover and amplify various MEV opportunities.


\vspace{-5pt}
\section{Countermeasures}
\label{sec:counter}

\noindent
\textbf{MEV Mitigation.}
Mitigating MEV removes the motivation for attackers to launch bribery attacks and maintains the security of the blockchain. 
Arbitrage and even MEV can be mitigated by eliminating the ability of the proposer to audit and order the transactions. 
Several works~\cite {kelkar2020order, kelkar2021themis, wei2024separation} propose order-fairness consensus algorithms to mitigate MEV. 
Some works~\cite{betarte2020towards, sasson2014zerocash} also propose to publish transaction contents only after the transaction is conﬁrmed to prevent auditing by block proposers. 
However, all these works require fundamental changes to the consensus layer. Existing systems take time to complete consensus protocol upgrades.

\vspace{3pt} 
\noindent
\textbf{Incentive Mechanism.}
We can also mitigate bribery attacks by improving the incentive mechanism. 
One of the most intuitive solutions is to increase the rewards for validators. 
That increases the cost of adversary for bribery attackers and raises the risk of bribery attackers. 
However, given that Ethereum wants to avoid inflation, more than such a simple incentive mechanism improvement is needed. 
Thus, we need further optimization. 

\vspace{-5pt}
\section{Conclusion}
This paper proposes BriDe Arbitrager, a bribery-enabled delayed and enhanced arbitrage tool for Ethereum 2.0.
It allows proposers controlling a limited fraction of voting powers to delay block production via bribery, and to increase their arbitrage profits via the DTOA algorithm.
Moreover, a bribery smart contract and a bribery client are proposed to provide trustless fairness and automation to the whole process.


By replaying Ethereum historical transactions, we estimate that BriDe Arbitrager is able to achieve an average 8.66\,ETH (16,442.23\,USD) daily revenue on the 50 liquidity pools that we selected through our heuristic liquidity pool selection strategy. 
Moreover, the arbitrage strategies discovered by BriDe Arbitrager have low initial capital requirements: most require capital less than 25\,ETH, which can be further reduced to less than 0.3\,ETH with flash loans. 
Compared to existing arbitrage detection algorithms, BriDe Arbitrager can boost profits by more than double. 
Note that the DTOA  algorithm could be enhanced to search for a wider range of MEV opportunities and to take full advantage of the delayed time introduced by BriDe Arbitrager, further improving profitability. 
This leaves an interesting direction for future work.

\bibliographystyle{IEEEtran}
\bibliography{main}





\vspace{-15pt}
\begin{IEEEbiography}[{\includegraphics[width=1in,height=1.25in,clip,keepaspectratio]{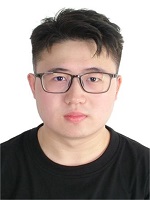}}]{Hulin Yang}
is currently a Ph.D candidate with the Department of Computer Science and Engineering, Southern University of Science and Technology. 
He received his B.E. degree in computer science and technology from Southern University of Science and Technology in 2021, and received his M.E. degree in electronic science and technology from Southern University of Science and Technology in 2024. 
His research interests are mainly in blockchain, decentralized finance and network economics.
\end{IEEEbiography}
\vspace{-10pt}
\begin{IEEEbiography}[{\includegraphics[width=1in,height=1.25in,clip,keepaspectratio]{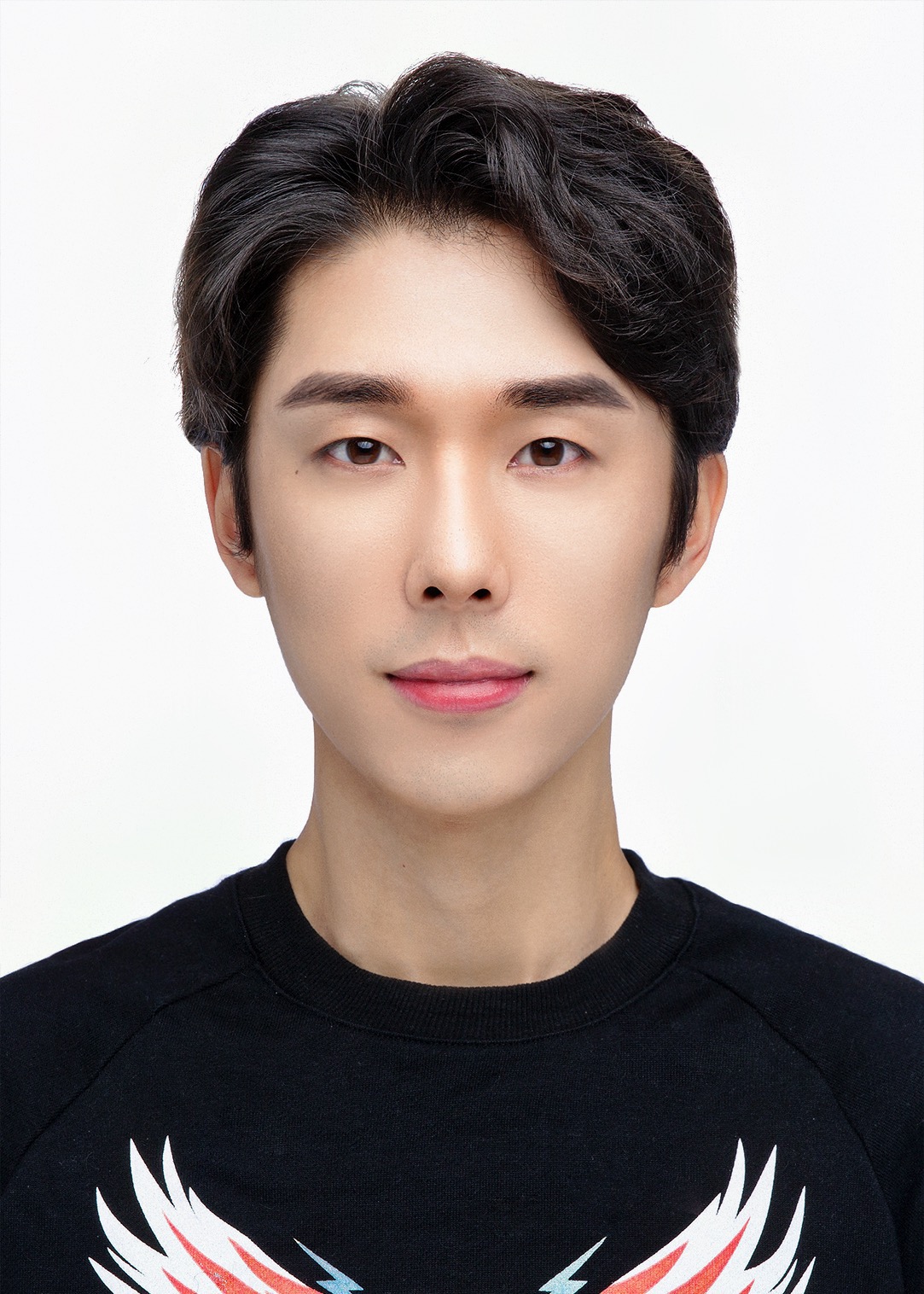}}]{Mingzhe Li}
is currently a Scientist with the Institute of High Performance Computing (IHPC), A*STAR, Singapore.
He received his Ph.D. degree from the Department of Computer Science and Engineering, Hong Kong University of Science and Technology in 2022.
Prior to that, he received his B.E. degree from Southern University of Science and Technology.
His research interests are mainly in blockchain sharding, consensus protocol, blockchain application, network economics, and crowdsourcing.
\end{IEEEbiography}
\vspace{-10pt}
\begin{IEEEbiography}
[{\includegraphics[width=1in,height=1.25in,clip,keepaspectratio]{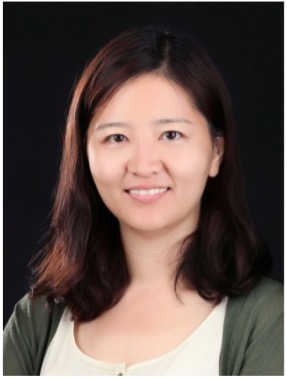}}]{Jin Zhang} 
is currently an associate professor with the Department of Computer Science and Engineering, Southern University of Science and Technology. 
She received her B.E. and M.E. degrees in electronic engineering from Tsinghua University in 2004 and 2006, respectively, and received her Ph.D. degree in computer science from Hong Kong University of Science and Technology in 2009. 
Her research interests are mainly in mobile healthcare and wearable computing, wireless communication and networks, network economics, cognitive radio networks and dynamic spectrum management. 
\end{IEEEbiography}
\vspace{-10pt}
\begin{IEEEbiography}[{\includegraphics[width=1in,height=1.25in,clip,keepaspectratio]{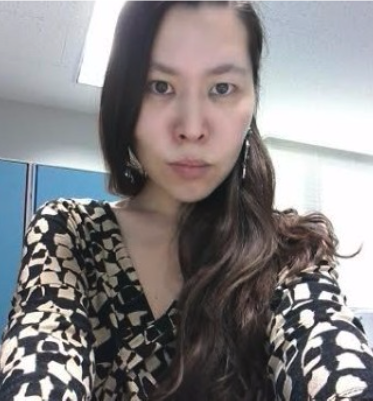}}]{Alia Asheralieva}
Alia Asheralieva obtained her Ph.D. degree from the University of Newcastle, Australia, in 2015. From 2015, she was with the Graduate School of Information Science and Technology, Hokkaido University, Japan. From 2017, she was with the Information Systems Technology and Design Pillar, Singapore University of Technology and Design. From 2018, she was with the Department of Computer Science and Engineering and Research Institute of Trustworthy Autonomous Systems of the Southern University of Science and Technology, China. She is currently with the School of Computer Science, Loughborough University, UK. Her research interests include edge computing, internet of things, blockchains, wireless networks and communications, unmanned aerial vehicles, game theory, combinatorial and stochastic optimization, machine learning and information security.
\end{IEEEbiography}
\vspace{-10pt}
\begin{IEEEbiography}[{\includegraphics[width=1in,height=1.25in,clip,keepaspectratio]{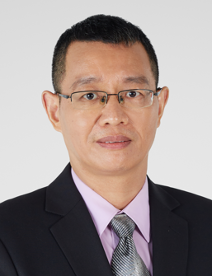}}]{Qingsong Wei}
received the PhD degree in computer science from the University of Electronic Science and Technologies of China, in 2004. He was with Tongji University as an assistant professor from 2004 to 2005. He is a Group Manager and principal scientist at the Institute of High Performance Computing, A*STAR, Singapore. His research interests include decentralized computing, Blockchain and federated learning. He is a senior member of the IEEE.
\end{IEEEbiography}
\vspace{-10pt}
\begin{IEEEbiography}[{\includegraphics[width=1in,height=1.25in,clip,keepaspectratio]{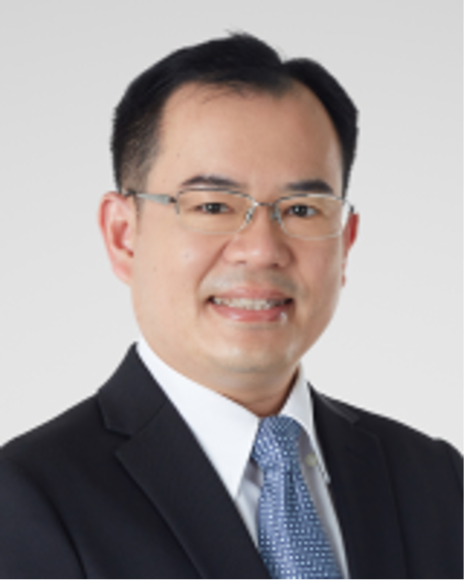}}]{Siow Mong Rick Goh}
received his Ph.D. degree in electrical and computer engineering from the National University of Singapore. He is the Director of the Computing and Intelligence (CI) Department, Institute of High Performance Computing, Agency for Science, Technology and Research, Singapore, where he leads a team of over 80 scientists in performing world-leading scientific research, developing technology to commercialization, and engaging and collaborating with industry. His current research interests include artificial intelligence, high-performance computing, blockchain, and federated learning.
\end{IEEEbiography}

\vfill

\end{document}